\documentclass[12pt]{article}
\usepackage{natbib}
\usepackage{graphicx}

\newcommand\ba{\begin{eqnarray}}
\newcommand\ea{\end{eqnarray}}
\newcommand\eq{\begin{equation}}
\newcommand\en{\end{equation}}
\newcommand\zt{\zeta_t(m_i,m_0,t_i,m,t)}
\begin{document}

\title{Source Mergers and Bubble Growth During Reionization}

\author{J.D. Cohn\thanks{jcohn@berkeley.edu}${}^{~ a,b}$ and
Tzu-Ching Chang\thanks{tchang@astron.berkeley.edu}${}^{~ b}$ \\ 
{\it ${}^a$Space Sciences Laboratory,}\\ 
{\it ${}^b$Theoretical Astrophysics Center, Dept. of Astronomy,}\\
{\it University of California, Berkeley, CA 94720}}
\maketitle
\begin{abstract}
The recently introduced models of reionization bubbles based on
extended Press-Schechter theory (Furlanetto, Zaldarriaga \& Hernquist
\cite{FZH04a}) are generalized to include mergers of ionization sources.
Sources with a recent major merger 
are taken to have enhanced photon production due to
star formation, and accretion onto a central black hole if a black hole
is present.
This produces a scatter in the number of ionized photons 
corresponding to
a halo of a given mass and a change in photon production over time
for any given halo mass.  By extending 
previous methods, photon production histories,
bubble distributions, and ionization histories are computed for several
different parameter and recombination assumptions.  The resulting
distributions interpolate between previously calculated limiting cases.
\end{abstract}

\section{Introduction and Background}
Reionization marks the historical event when hydrogen in the universe
transformed from mostly neutral to mostly ionized; it has only recently
become accessible via observations and numerical simulations and has
stirred great interest and advances in our understanding of early
structure formation (see e.g. reviews by Barkana \& Loeb \cite{BarLoe01},
Loeb \& Barkana \cite{LoeBar01}, Cooray \& Barton \cite{CooBar05},
Loeb \cite{Loe06}).

Current observations suggest that the process of reionization is
complex, perhaps lasting over an extended period of time (for a 
review, see Fan, Carilli \& Keating \cite{FanCarKea06}).  
The optical depth to electron scattering seen by WMAP
(Spergel et al \cite{WMAP}) can be modeled with reionization starting by
$z \geq 10$.  On the other hand, spectroscopic observations
of high-redshift quasars find Gunn-Peterson absorption due to
intervening neutral hydrogen in the intergalactic medium (Becker et al
\cite{Bec01}, Fan et al \cite{Fan03},White et al 
\cite{Whi05}), although with significant fluctuations along different 
lines of sight (Oh \& Furlanetto \cite{OhFur05}), which suggests that 
the tail end of reionization occurs at $z \sim 6$.

Theoretically, questions about reionization range from the fundamental
timeline of reionization to detailed characteristics.  The
latter include the nature of
ionizing sources, the topology of the ionized regions and their
evolution, and the behavior of the ionized regions relative to matter
overdensities and voids.  Numerical approaches (e.g. Gnedin
\cite{Gne00}, Razoumov et al \cite{Raz02}, Ciardi, Stoehr \& White
\cite{CiaStoWhi03}, Sokasian et al \cite{Sok03, Sok04}) are fruitful
due to the complex nature of the sources and nonlinear clustering
involved.  For small regions, simulations can track the nonlinear
effects, although much known physics is too difficult to incorporate,
and much unknown physics remains.  However, the large disparity in
size between the ionized regions (which can grow to be tens of Mpc
comoving, e.g.  Wyithe \& Loeb \cite{WyiLoe04}, Furlanetto, 
Zaldarriaga \& Hernquist \cite{FZH04a}) 
and the small scale distributions and physics of
sources producing the photons is difficult to capture in a simulation.
Recent numerical simulations have just begun to combine large and
small scale effects in larger volumes (see for instance Kohler, Gnedin
\& Hamilton \cite{KohGneHam05}, Iliev et al \cite{Ili05}).

Analytic models are useful to encompass the wide range of scales in
order to try to identify generic behavior.  In addition, analytic
models allow exploring a larger range of parameters and other assumptions.
This paper extends an analytic model introduced by Furlanetto,
Zaldariagga and Hernquist \cite{FZH04a} (hereafter FZH) to
characterize the growth of ionized regions or bubbles.  The basic idea
is to move beyond the initial step of considering HII regions around
individual collapsed halos (e.g. Arons \& Wingert \cite{AroWin72},
Barkana \& Loeb \cite{BarLoe01}) to consider regions around several
collapsed halos.  The simplest form goes as follows: collapsed halos
of mass $m$ are able to ionize a mass $M$ where \eq
\label{ionave}
M = \zeta m
\end{equation}
and $\zeta$ is a constant. 
Halos need to be massive enough to produce photons,
thus $m>m_{min}(z) = m(T=10^4 K,z)$ is imposed, i.e. they need to be massive
enough to have efficient atomic Hydrogen line cooling.  
For a constant $\zeta$, independent of $m$, the size of ionized regions
can be calculated immediately in closed form: an ionized region of mass $M$ has
the fraction of mass $f_{coll}$ bound in halos
above $m_{min}(z)$  equal to or greater than $\zeta^{-1}$.
Using extended Press-Schechter theory (references below), the mass
fraction $f_{coll}$ in a fully ionized bubble of mass $M$
then obeys
\eq
\label{fconst}
f_{coll} = {\rm erfc} [\frac{\delta_c(t)-\delta_{M}}{
\sqrt{2(\sigma_{min}^2 - \sigma^2(M))}}] > \zeta^{-1} \; .
\end{equation}
This constrains the average overdensity $\delta_M$ for any $M$.
Here $\sigma_{min}$ denotes the density fluctuations smoothed on a scale of
mass $m_{min}(z)$, likewise for $\sigma^2(M)$, and
$\delta_c(t)$ is the threshhold density for collapse.
The overdensity required for collapse is $\delta_c =1.686$,
in this picture the densities stay constant and the threshold lowers
with time, so $\delta_c(t) = \delta_c /D(z(t))$ where $D(z)$ is the growth
factor.  
The bubble's
average overdensity relative to $\delta_c(t)$ is $\delta_M$, corresponding to
physical overdensity $\delta_{R,M}(t) = \delta_M D(z(t))$, and
$V_M (1+\delta_{R,M}(t)) \bar{\rho} = M$, where $V_M$ is
the bubble volume and $\bar{\rho}$ is the mean density.  
The condition for a bubble to be ionized, 
equation \ref{fconst}, can be rewritten in a form
that is more easily generalizable,
\eq
\label{zetaf}
\zeta f_{coll} = 1  \; .
\end{equation}
The function $\delta_M$ allows one to determine the number distribution of
bubbles of mass $M$ and other properties discussed below.

In FZH, recombinations were included implicitly in $\zeta$.  A more
sophisticated approach balancing the total number of ionizations and
recombinations per unit time in the intergalactic medium (IGM) was
introduced in Furlanetto \& Oh \cite{FurOh05} (hereafter FO05), where
they considered two models of IGM gas density distribution for
recombinations: one model adopted an analytical fit of gas density
to simulations at low redshift by Miralda-Escude,
Haehnelt \& Rees \cite{MirHaeRee00}, and extrapolated it to high z;  a
second model considered minihalos as photon sinks. Minihalos are neutral, 
dense blobs with mass $10^6 M_{\odot}$, randomly distributed in the IGM.  

These results were extended by Furlanetto, McQuinn \&
Hernquist \cite{FurMcQHer05} (hereafter FMH05) and others.  For example,
FMH05 include the effects of different halo mass functions,
a mass dependent $\zeta(m)$, and stochastic fluctuations in the galaxy
distribution.  They found the consequences of these additional effects
on the bubble size distributions and corresponding evolution in time,
the emissivity inside a bubble, the observable neutral hydrogen 21 cm power
spectra, and the kinetic Sunyaev-Zel'dovich 
signal.  Tools for calculations of the latter were
developed by Zahn et al \cite{Zah05} and McQuinn et al \cite{McQ05},
they also extended many aspects of the approach.  Other consequences
of this model for observables have been calculated, for example in
Furlanetto, Hernquist \& Zaldarriaga \cite{FurHerZal04}, Furlanetto,
Zaldarriaga \& Hernquist \cite{FZH04b,FZH06}, McQuinn et al
\cite{McQ05b} and Alvarez et al \cite{Alv05}.

Here we extend this model further to include the effects of
major mergers.
The original models are based on a time
independent $\zeta$ which only depends upon mass. 
Halos which have recently had
a major merger produce an increased number of photons, due to starbursts
(1:3 halo mass ratios)
or induced black hole accretion (1:10 halo mass ratios). Consequently 
major mergers
introduce a time dependence in the average of $\zeta$ and a
scatter in $\zeta$.
We expect halo major mergers to be important, as even
the smallest objects producing photons (chosen to be those with $T >
10^4 K$) have masses several orders of magnitude above $M_*$ at high
redshifts, and thus are highly biased and very actively merging.\footnote{
$M_*$, a function of time,
is the mass at which $\sigma(M)= \delta_c(t)$. An estimate of the
formation minus destruction rate for halos of mass $M$ is proportional
to
$\delta_c(t)/\sigma^2(M) - 1/\delta_c(t)$, see 
e.g. Kitayama \& Suto \cite{KitSut96}.  For $M>M_*$ this rapidly increases
with $M$ and can be thought of as indicative of not only a high formation 
rate but also a high major merger rate--more detailed calculations
confirm this trend.}
We use estimates of photon production due to quiescent
star formation, merger induced starbursts and black hole accretion.
As the photon production estimates have many unknowns, we are especially
interested in features which characterize the merging itself rather than the
details we choose.\footnote{ Ideally this could eventually be turned
  around to constrain assumptions going into the 
photon production estimates but it is not clear how
  constraining that might be given the vast uncertainties.}

There are several related studies in this active area of research.
This is not the mergers of the bubbles,
which was already considered in FO05.  The scatter discussed is not 
the scatter
(Furlanetto et al \cite{FurMcQHer05}) in $\zeta$
due to the stochasticity of the number of collapsed halos (Barkana \& Loeb 
\cite{BarLoe04a}, Babich \& Loeb \cite{BabLoe05}), or the difference
between metal rich and metal poor regions (Barkana \& Loeb \cite{BarLoe05}).
Other related work includes the calculation of the effects of mergers
in producing cumulative number of photons by quasars up to redshifts 
$\sim 5$, e.g. by
Wyithe \& Loeb \cite{WyiLoe02} and Madau et al \cite{Mad04}.  Here we
want the effects of the mergers in producing a time dependent photon
production rate, with scatter, for collapsed halos of a given mass,
and the consequences for the enclosing bubbles.

Section \S2 gives the analytic framework and formulae for the merger rates,
recombinations and scatter.
The natural quantity to use is the instantaneous photon production rate 
which then
determines the time derivative of $\zeta_t$ rather than $\zeta$ itself.
The resulting formalism
is a natural extension of earlier work
(e.g. FZH04, FO05, FMH05): their early time limit took reionization to
only depend on the halos present at the time under consideration
and included recombinations implicitly,
the late time limit imposed equilibrium between
instantaneous photon production and recombination rates.  The extension of
these earlier formalisms, introduced here, finds the total
number of ions present by integrating the photon production
rate over time (due to the halos present at any instant), 
and subtracting recombinations.
Three prescriptions 
for recombinations are used.  The scatter requires additional assumptions,
we describe and choose a simple case next.
In section \S3,
we calculate the analytic form of the time derivative $\zeta_t$ in
terms of the star formation and black hole accretion parameters and then
explicitly show its dependence upon mass and redshift for a few
examples. 
Section \S4  shows the effects on bubbles using a fiducial model and
some variations:
the bubble overdensity $\delta_M$ as a function of 
bubble mass $M$ (as a function of redshift and initial conditions),
the characteristic
bubble sizes and ionization fractions over time $\bar{x}_i$,
and the bubble size
distributions associated with the (major) merger scatter.
Section \S5 concludes.  An appendix spells out the basic Extended Press-
Schechter quantities, details the estimates used to
find the minimum halo mass for a black hole to be present and
describes some other possible assumptions for scatter.
The term ionized mass fraction refers to the mass found in ionized
bubbles unless otherwise stated.

\section{Analytic Framework}

In this section we describe the analytic calculations, assumptions
and formulae used in the rest of the paper: the halo mass function,
major merger rates, recombinations and the merger related scatter.
The input parameters to most of our calculations are
$\Omega_m = 0.3, \Omega_\Lambda = 0.7, \sigma_8 = 0.9, \Omega_bh^2 = 0.02,
h = 0.7$
a scale invariant initial power spectrum $n=1$, and the Eisenstein-Hu
\cite{EisHu97} transfer function.  Our fiducial model has these parameters.
 We considered several changes of
cosmology.  We used the above and only changed $n$ to 1.05,
which enhances fluctuations for large $k$, i.e.
small scales, and we fixed to the above but instead took $\Omega_bh^2 = 0.225$.
Another variant we considered was the set of parameters from the
recent WMAP analysis (Spergel et al \cite{WMAP}): $\Omega_m = 0.24, 
\Omega_\Lambda = 0.7, \sigma_8 = 0.74,\Omega_bh^2 = 0.0223,h = 0.73$, we discuss
these at the end.
Masses are defined in terms of $h^{-1} M_\odot$ and all
lengths are comoving.

\subsection{Halo Mass Function}
The starting point for the extended 
Press-Schechter bubble model of reionization 
is the number density of collapsed halos of dark matter.
We use the Press-Schechter formalism \cite{PS} to estimate 
numbers and source major merger rates.   Examples of the resulting number
densities and properties have been detailed by  
Barkana \& Loeb \cite{BarLoe01} and Mo \& White \cite{MoWhi02}.

Our use of the Press-Schechter mass function warrants some discussion.
The Sheth-Tormen \cite{SheTor02} mass function
is a better fit to simulations at low redshifts; mass functions found
in high redshift simulations (Jang-Condell \& Hernquist
\cite{JanHer01}, Reed et al \cite{Ree03}, Heitmann et al \cite{Hei06},
Iliev et al \cite{Ili05}) often lie between Sheth-Tormen and
Press-Schechter (the best agreement for Heitmann et al was with the
fitting function of Warren et al \cite{War05}).  Agreement with dark
matter simulations is not necessarily indicative of appropriateness
either, as baryons and dark matter are not as tightly coupled at high
redshifts (e.g.  Naoz \& Barkana \cite{NaoBar05}, Bagla \& Prayad
\cite{BagPra06}).  For the original model (which we modify here), the
the effect of using the Sheth-Tormen mass function instead of the
Press-Schechter mass function for the number of sources was seen to be
small (Furlanetto et al \cite{FurMcQHer05}).  We altered the ``tilt''
$n$ in one set of runs from $n=1$ to $n=1.05$ to see the effect of
increased small scale structure as found in these simulations.

The Press-Schechter mass function is the most tractable and
our interest is in trends rather than exact numbers.  With it,
the number of recently merged halos can be calculated easily using
extended Press Schechter theory
(Bond et al \cite{BCEK}, Lacey \& Cole \cite{LacCol93,LacCol94},
Bower\cite{Bow91}, Kitayama \& Suto \cite{KitSut96}).\footnote{These
are actually formulae for fast mass gain and can include accretion.
Analytic formulae differentiating between major mergers and accretion have
been found by Raig et al \cite{Rai98},
Salvador-Sole et al \cite{Sal98} and Andreu et al \cite{And01}.}
There are
problems as well 
with extended Press-Schechter theory, especially when considering
large mass ratios in merger rates
(see Benson et al \cite{BenKamHas05} for a recent discussion of this).
We restrict ourselves to major mergers, with small mass ratios.
In addition to being in the regime where extended Press-Schechter works the
best, this allows predecessors and final halos to be identified uniquely
(a halo only has one predecessor with mass greater than half the final
halo mass).

\subsection{Photon Sources}
\subsubsection{Major Merger Rates}
After a major merger, a halo produces extra photons for some specified
amount of time, denoted $t_{ion}$.  We are interested in the number of
halos of a given mass which are still ``active,'' i.e. which are still
producing extra photons due to a recent major merger.  We ignore the lag
between the actual merger and the beginning of the photon production,
which should not change the general trends of interest.  We also
assume that the extra photon production rate is constant during
$t_{ion}$ after the merger, so that we do not care about when exactly
the merger happened, only that it happened within this relaxation
time.  

The number of recently merged halos with mass $m$ at time $t$
is the product of two factors, depending upon two times: the time of
the merger $t_i$ and the subsequent time of observation $t$, where the
latter is close enough to $t_i$ for the halo to still be excited from
the merger, i.e. $t - t_i < t_{ion}$.
The first factor is the fraction of halos that have mass $m_0$ which have
jumped at time $t_i$ from mass
$m_i$, $\dot{P}_1(m_i\to m_0;t_i)  \frac{m_0}{m_i} dm_i dt_i$.  (The overdot
denotes derivative with respect to $t_i$.)
The factor of $m_0/m_i$ makes the conversion from
the number of points coming from $m_i$ halos (the quantity given by the
formalism) to the total number of points in their descendant $m_0$ 
halos.\footnote{If the $m_0$ halo has
come from an $m_i$ halo, all of its mass was not previously in
the $m_i$ halo, only $m_i/m_0$ of its mass.  But the quantity wanted
here is the full mass of halos which have had mergers, i.e. halos which
now have mass $m_0$ but had such an $m_i$ component earlier; thus the 
factor $m_0/m_i$ is included.}
We then want this "excited" halo of mass $m_0$ to be in a halo of 
mass $m$ at time $t$.
The probability that a point in a halo of mass $m$ at time $t$ was in
a halo of mass $m_0$ at time $t_i$ is $P_1(m_0,t_i|m,t)dm_0$.  Formulae
for $P_1$ and its derivative are in the appendix.

For bubbles we need to go one step further and consider the regions of mass
$M$ surrounding these mass $m$ sources.  The probability that
a halo (bubble) of mass $M$ and overdensity $\delta_M$ contains a halo of mass
$m$ at time $t$ is denoted by
$P_1(m,t|M,\delta_M)dm$, corresponding to a number density of halos
$n(m,t|M,\delta_M) dm = \frac{\bar{\rho}}{m} P_1(m,t|M,\delta_M)dm$.
(When a time $t$ rather than $\delta$ 
is written as an argument of $P_1$ the implicit assumption is
that one uses $\delta_c(t)$ in $P_1$, the threshold density corresponding to
time $t$.)  
Thus the number of halos which have merged at time $t_i$ from mass $m_i$ to
mass $m_0$ and are in a halo of mass $m$ at time $t$ within a bigger halo
(bubble) of mass $M$ and overdensity $\delta_M$ is given 
by (see Appendix for more discussion and details):
\eq
\label{mergeprob}
V_M  n(m,t|M,\delta_M) 
\dot{P}_1(m_i \to m_0,t_i) P_1(m_0,t_i|m,t) 
\frac{m}{m_i} dm dt_i dm_i dm_0 \; .
\end{equation}

\subsubsection{Total Photon Rates}
The rate photons are produced in a region with mass $M$ and 
overdensity $\delta_M$ is
the number of additional photons due to the mergers, plus
the quiescent number of photons due to the collapsed halos present
even if no merger has occurred.  (We assume that each photon produced
corresponds to an ionization in the region, in principle the bubble region
can have
some ``escape fraction'' which would effectively lower the number of
ionizations in a region relative to the number of photons produced.)  
A constant $\zeta$ for sources of lifetime
$\Delta T$ gives a $\zeta_t = \zeta/\Delta T$.  More generally,
there is a dependence upon the mass of the initial and final masses before
the merger ($m_i,m_0$), the mass $m$ at the time of interest, and $t_i,t$ 
to give $\zt$.  The quiescent rate 
is taken to be $\zeta_{t,q}(m)$.  These quiescent photons are from stars
not created during starbursts, i.e. ongoing star formation, in part due
to accretion.  Putting this together to get the rate
mass is ionized gives

\eq
\label{photonrate}
\begin{array}{l}
\int dm n(m,t|M,\delta_M)   \frac{m}{\bar{\rho}} \\
\{
\int dt_i dm_i dm_0 \zt
\dot{P}_1(m_i \to m_0,t_i) P_1(m_0,t_i|m,t) 
\frac{m}{m_i}  + \zeta_{t,q}(m) \} \\
= \int dm n(m,t|M,\delta_M) \frac{m}{\bar{\rho}} \zeta_{t,a}(m)
\\
=  (\zeta f)_t 
\end{array}
\end{equation}

The above defines  $(\zeta f)_t$ and $\zeta_{t,a}(m)$.
Section \S3 derives estimates for $\zt$ using models for
star formation and black hole accretion, and gives parameters and
limits for the integrals.  In the integrals, 
the mass limits are determined by the mass ratio
criteria for major mergers.  In all cases
we require $m_{i,max}/m_0 >m_i/m_0 > 0.5$.  The lower limit 
means some major mergers are neglected, but ensures that
no major merger is counted twice, as only one halo with $m_i > m_0/2$
can end up in $m_0$.  This is thus a lower limit on the number of mergers.
The time limits in the integrals are determined
by the time scales of relaxation after the mergers.  Both the
mass and time limits can differ for starbursts and black hole accretion.

In principle $m_{0,min}$ can be as small 
as $m(T=10^4 K,z) =  \frac{2.04 \times 10^9}{(1+z)^{3/2}} h^{-1} M_\odot$.
However the probabilities above double count a halo which has 
two mergers within a relaxation time, which is more likely if $m_{0,min}$ is
chosen to be small.  For the number of photons contributed, this is 
perhaps accurate (but perhaps not, e.g. for starbursts
there might be gas missing after the first merger for some period of time).
However for the scatter it will
give two final halos with recent mergers rather than just the one which
had two mergers.  However, if $m_{0,min}$ is taken to be too large,
the number of photons will be undercounted and the scatter 
underestimated. 
 We used both $m_{0,min} = 
0.1 m$ and $m_{0,min}= m_{min}(T=10^4K,z)$ in our calculations,\footnote{If 
we take the limits of $m_0$ to be independent of $m$, then
the integral over $m$ can be done
explicitly using an identity for composing the probabilities.  This 
gives the mass fraction which had
mergers at $t_i$ to mass $m_0$ in a halo of mass $M$ as
\begin{eqnarray}
\int dm
\dot{P}_1(m_i\to m_0;t_i)
 \frac{m}{m_i} P_1(m_0,t_i|m,t) P_1(m,t|M,\delta_M) dm_i dm_0 dt_i \\
\; \; \; \; \; \; \; =
\dot{P}_1(m_i\to m_0;t_i)
\frac{m}{m_i} P_1(m_0,t_i|M,\delta_M) dm_i dm_0 dt_i \\
\end{eqnarray}
i.e. it doesn't matter what $m$ halo the original $m_0$ halo is in at
time $t$, just that it is in the bubble $M$. }
for nontrivial mass dependence of star formation
(referred to as $\alpha = 2/3$ below) these two choices are essentially
indistinguishable.

The ionized mass fraction
in a region of mass $M$ with overdensity $\delta_M$,
the generalization of equation \ref{zetaf}, is then
\eq
\label{fullzet}
\int dt [(\zeta f)_t  - R_{recomb}(R_{ion}(M),\delta_M,t)] \; .
\end{equation}
where $R_{recomb}(t)$ is the instantaneous recombination rate in
the bubble volume, depending upon $R_{ion}(t)$,
the effective radius of the ionized region at this time.  The dependence
upon $R_{ion}$ takes into account that not all regions inside a bubble
at time $t$ are ionized at earlier times: recombinations
can only occur where there are ions already present.  As the first
term gives the total number of photons produced in the region, assumed
to produce the same number of ions, recombinations can be directly
subtracted off. 

\subsection{Recombination}
Recombinations decrease the number of ions present.
We use the
two estimates for the recombination rate for an
ionized region of radius $R_{ion}(t)$ and overdensity $\delta_M$ chosen
by FO05 and a third by Mellima et al \cite{Mel06}.
The recombination rate per hydrogen atom
inside an ionized region of radius $R_{ion}$ at
time $t$ and overdensity $\delta_{M}$ is
\eq
\alpha_A(T) {n}_e C(R_{ion}(t)) =
A_u (1+\delta_{R,M}(t))C(R_{ion}(t)) \; ,
\end{equation}  
where $n_e = \bar{n}_e (1+\delta_{R,M}(t))$.  The physical overdensity
$\delta_{R,M}(t)$ enhances
the rate at mean density $A_u = \alpha_A(T) \bar{n}_e  = 2.4 
\times 10^6/Myr$.  (We  use  $\alpha_A(T= 10^4K)$ for case A recombination,
case B recombination (optically thick) takes $A_u \rightarrow 0.6 A_u$.)
The three recombination models have different clumping factors, described
in subsections below.

To go from the ionization rates per hydrogen atom in the bubble to the change 
in ionization
fraction in equation \ref{fullzet} requires the comparison of 
counts\footnote{We thank S. Furlanetto for discussions about this.}: counts
of ions present and counts of recombinations depleting ions.
The ionized mass fraction times the hydrogen mass of the bubble gives the number
of ions in the bubble at a given time.  The recombination rate at the
same time is the rate per hydrogen atom times the number of ionized hydrogen
atoms, $\bar{n}_h (1+\delta_{R,M}(t)) V_{ion}$.
Here $V_{ion}$ is the volume of the region where the hydrogen is ionized
and $M_{ion}$ is its mass.
Dividing by the total mass of the bubble and
noting $V_{ion} = \frac{M_{ion}}{M} \frac{M}{(1+\delta_{R,M}(t)) \bar{\rho}}$,
the resulting ionization fraction with recombinations included,
at some time $t$, is  
\eq
\label{fullzetr}
\int^t dt' [(\zeta f)_{t'}  - A_u \frac{M_{ion}(t')}{M} (1+ \delta_{R,M}(t'))
C(R_{ion}(t')) ] \; . 
\end{equation}
The mechanics of doing this integral are discussed in section \S4, we
now describe the clumping factors $C(R_{ion})$ for the three different
recombination models.

\subsubsection{The MHR model}

The first model (hereafter called the MHR model) 
is a smooth IGM gas distribution of 
Miralda-Escude, Haehnelt \& Rees \cite{MirHaeRee00}.
They use the volume density distribution of IGM gas, $P_V(\Delta)$
($\Delta = \rho/\bar{\rho}$) fit to simulations at $z \sim 2-4$, finding
\eq 
P_V(\Delta) d\Delta = A_0 \Delta^{-\beta}
\exp[-\frac{(\Delta^{-2/3} - C_0)^2}{2(2 \delta_0/3)^2}] d\Delta
\end{equation}
with $\beta = 2.5$. $A_0$ and $C_0$ are set by requiring mass 
and volume normalization and $\delta_0$ is the variance of
density fluctuations smoothed on the Jeans scale for an ionized medium
(at higher $z$, $\delta_0 = 7.61/(1+z)$ to better than 1\%).  The distribution
$P_V(\Delta)$ is taken to be
independent of environment (rather than depending upon $\delta_M$).
For recombination, they assume all gas below some density threshold
$\Delta < \Delta_i$ is ionized and
everything above is shielded.   
One finds $\Delta_i$ by noting that recombination limits bubble growth--
i.e. the mean free path 
 $\lambda_i = \lambda_0[1-F_V(\Delta_i)]^{-2/3}$ is the radius
of the ionized region $R_{ion}$.  Here
$\lambda_0 H(z) = 60 {\rm km \; s}^{-1}$ (in physical units)
and $F_V(\Delta_i)$ is the fraction of volume with $\Delta < \Delta_i$.
(If the region's radius $R_{ion}< \lambda_0$ then the assumption is that
recombination is negligible and $\Delta_i$ is set to zero.)
The factor $C$ is then 
\eq
\label{eq:cmhr}
C_{MHR}(R_{ion}) = \int_0^{\Delta_i} d \Delta P_V(\Delta) \Delta^2 \; ,
\end{equation}
again, note
$R_{ion}$ enters implicitly via $\Delta_i(R_{ion})$.
At large $R_{ion}$ $C_{MHR}(R_{ion})$ 
tends to asymptote to a constant, and for most (all) radii
it is smaller than $C_{mh}(R_{ion})$ ($C_{MIPS}$) described below.

\subsubsection{Minihalos}

The second model is a minihalo model (FO05) of small dense absorbing clumps,
taken to have mass of $M_{mh} =  10^6  M_\odot$ 
with comoving mean free path
\eq
\begin{array}{ll}
\ell_{mh} &= \frac{1}{\pi n_{mh} R_{mh}^2} \\
&\sim 15.7 h (\frac{M_{mh}}{ 10^6 M_\odot})^{1/3}
(\frac{0.05}{f_{mh}})
(\frac{\Delta_{mh}}{18 \pi^2})^{2/3} (\frac{\Omega_m h^2}{0.15})^{-1/3}
{\rm h^{-1}\; Mpc}
\end{array}
\end{equation}
and
\eq
\label{eq:mini}
C_{mh}(R_{ion}) =(1-f_{mh})^2 \exp(R_{ion}/\ell_{mh})
\end{equation}
where $f_{mh} = 0.05$ is the mass fraction taken to be in minihalos.
Unlike the other two models it is redshift independent.  Note that the
effect of minihalos on recombination and thus reionization is likely
to be more complicated than this simple prescription, which is
probably an overestimate (Iliev, Scannapieco \& Shapiro
\cite{IliScaSha05}, Ciardi et al \cite{Cia06}).
 
\subsubsection{The MIPS model}
A third prescription from clumping, matched to numerical simulations
is by Mellema, Iliev, Pen \& Shapiro \cite{Mel06} (MIPS), which is an
improved fit for the IGM clumping factor to the one given by Iliev et
al \cite{IliScaSha05}.  They used a $N$-body simulation with a 3.5
$h^{-1}$ Mpc box and $3248^3$ particles to get a mass resolution down
to the Jeans mass; they then took out all found halos, including
minihalos, to eliminate their contributions to the IGM density field.
The resulting IGM was fit to give an $R_{ion}$ independent clumping
factor: \eq
\label{eq:mips}
C_{MIPS} = 27.466 \exp(-0.114z + 0.001328 z^2)
\end{equation}
It resembles the MHR clumping factor at large $R_{ion}$ as the
former asymptotes to a constant for large $R_{ion}$ and high $z$,
but is larger numerically, and thus more effective
at suppressing bubble formation than the MHR model.
This will be referred to as the MIPS model in the following.

\subsection{Scatter}
\label{scatter}
Extended Press-Schechter gives the fraction of halos which have had recent
mergers.  This is an average quantity, just as the number of
collapsed halos in a larger region of mass $M$ is given only
on the average by $n(m,t|M, \delta_M)$.  In any region a scatter in the
number of sources will lead to a scatter the number of ionizing
photons and thus in bubble sizes.
For $n(m,t|M,\delta_M)$,
FMH05 took the scatter due to the stochastic nature of the collapsed
halo distribution (halos of mass $m$) in the larger bubbles of mass $M$
to be Poisson.  
Numerical simulations have found this scatter to be within a factor of
two of Poisson for the cases studied by Sheth \& Lemson \cite{SheLem99}
and Casa-Miranda et al \cite{Cas02}.  In equations, the ionized mass
fraction is
\eq
\int dm \zeta(m) \frac{m}{\bar{\rho}} n(m,t|M,\delta_M)
\end{equation}
and they assumed for the scatter
\eq
\begin{array}{ll}
\langle n(m,t|M,\delta_M) n(m',t|M,\delta_M) \rangle &=
n(m,t|M,\delta_M) n(m',t|M,\delta_M) \\
&+
\frac{\delta_D(m-m')}{V_M} n(m,t|M,\delta_M) \; .
\end{array}
\end{equation}
where $\delta_D$ is a Dirac delta function.

For our case we want the scatter of
\eq 
\int dt \int dm n(m,t|M,\delta_M) \frac{m}{\bar{\rho}} \zeta_{t,a}(m) \; ;
\end{equation}
i.e.
\eq
\int dt dt' dm dm' \frac{m m'}{\bar{\rho}^2}
\langle
n(m,t|M,\delta_M) \zeta_{t,a}(m) n(m',t'|M,\delta_M) \zeta_{t',a}(m')
\rangle \; ;
\end{equation}
Thus instead of the scatter of
$n(m,t|M,\delta_M)$ for fixed time above, we need
\eq
\langle n(m,t|M,\delta_M) n(m',t|M,\delta_M) \rangle \; .
\end{equation}
In addition, $\zeta_{t,a}(m)$ has scatter.
To our knowledge the
appropriate distributions have not been calculated in
numerical simulations, and thus additional assumptions are
required to continue.

The simplest assumption is to take the distribution of
$n(m,t|M,\delta_M)$ to be Poisson
both in mass and in time:
\eq
\begin{array}{ll}
\langle n(m,t|M,\delta_M) n(m',t'|M,\delta_M)\rangle
&=  n(m,t|M,\delta_M) n(m',t'|M,\delta_M)
\\
& +\frac{1}{V_M}\delta_D(m-m') \delta_D(\delta(t)-\delta(t'))
n(m,t|M,\delta_M)
\end{array}
\end{equation}
Note that $n$ is a function of  $t$ only through $\delta(t)$ and
thus $\delta(t)$ is taken to be the argument of the Dirac delta function.
Using $\delta_D(t-t')$ would give the wrong dimensions. 

The second assumption is for the scatter of
\eq
\zeta_{t,a}(m) = \int dt_i dm_i dm_0 \zt
\dot{P}_1(m_i \to m_0,t_i) P_1(m_0,t_i|m,t) 
\frac{m}{m_i}  + \zeta_{t,q}(m)  \; .
\end{equation}
We define $n^a(m_0|m,t)$ which appears in  
$\zeta_{t,a}(m)$ the following way:
\eq
\int dt_i dm_i \zt  \dot{P}_1(m_i \to m_0,t_i) \frac{m_0}{m_i} P_1(m_0,t_i|m,t) = 
\sum_{a=1}^3 \frac{m_0}{\bar{\rho}} n^a(m_0|m,t) \zeta_t^a (m_0) \; .
\end{equation}
The number density $n^a(m_0|m,t)$
counts the recently merged halos of mass $m_0$  in a halo of mass $m$,
with the index $a$ denoting the range of integration for $m_i,t_i$.
There are three ranges, corresponding to starbursts, 
black hole accretion or both.
For each $a$, $\zt$ only depends on which of these three
cases is at hand and the values of $m_0,m$, and so it is denoted
$\zeta_t^a(m_0)$.  Values for $ m\zeta_t^a(m_0)$ are
derived in the section \ref{sec:recipe} and displayed in Table 1 for
the starburst and black hole accretion case (the sum of these is needed
when both are present).
The Poisson assumption for $n^a(m_0|m,t)$ is then
\eq
\langle n^a(m_0|m,t) n^b(m_0'|m,t) \rangle = \frac{\delta(m_0-m_0')}{V_m}
 \delta^{ab} n^a(m_0|m,t)
\end{equation}
There is also an assumption being made that the number with scatter is
related to the product of $\dot{P}_1 P_1$, rather than a separate scatter
for the merger rate to mass $m_0$ and then scatter
for the inclusion of these $m_0$ halos in $m$.

The full scatter is then
\eq
\begin{array}{l}
\langle \zeta_{t,a}(m) n(m,t|M,\delta_M)
\zeta_{t',a}(m') n(m',t'|M,\delta_M)
\rangle \\
\; \; = \zeta_{t,a}(m) n(m,t|M,\delta_M)
\zeta_{t',a}(m') n(m',t'|M,\delta_M)
\\
\; \; +\frac{1}{V_M}\delta_D(m-m') \delta_D(\delta(t)-\delta(t'))
n(m,t|M,\delta_M)
\langle\zeta_{t,a}^2(m) \rangle \\
\; \; = 
 \zeta_{t,a}(m) n(m,t|M,\delta_M)
\zeta_{t',a}(m') n(m',t'|M,\delta_M)
\\
\; \; + \frac{1}{V_M}\delta_D(m-m') \delta_D(\delta(t)-\delta(t'))
n(m,t|M,\delta_M) \{ \zeta_{t,a}^2(m) \\
\; \;+ 
\int dm_0 dm_i dt_i  (1+\delta_R(t,t_i))
(\zt m)^2 \frac{1}{m m_i}\dot{P}_1(m_i \to m_0,t_i)
P_1(m_0,t_i|m,t) \}
\end{array}
\end{equation}
Here $m= (1+\delta_R(t,t_i))\bar{\rho} V_m$ was used,
where $\delta_R(t,t_i)$ 
is the physical overdensity $\delta(t)D(z(t_i))$
and the combination $\zt m$ was pulled out.  

We then define the scatter in $\int dt (\zeta f)_t$ via
\eq
\label{eq:scatter}
\Delta (\int (\zeta f)_t dt)^2 =
\langle \int dt dt' (\zeta f)_t (\zeta f)_{t'} \rangle
- \langle \int dt  (\zeta f)_t \rangle \langle \int dt'(\zeta f)_{t'} \rangle
\end{equation}
We calculated the
consequences of this simplest set of assumptions for our models;
some other possibilities are described in the appendix.

The scatter is an integral over time.  In practice we cannot
integrate back to arbitrarily early times,
and so an initial condition and initial scatter are needed.
See below for discussion.

\section{Photon increase from mergers}
The analytic expressions above require estimates of $\zt$  and
the quiescent rate $\zeta_{t,q}(m)$ to give concrete results.
The two main sources of extra photons due to a major merger are
starbursts and, if a central black hole is present,
accretion onto a central black hole (Kauffmann \& Haehnelt
\cite{KauHae00}, Cavaliere \& Vittorini \cite{CavVit00}).
Merger induced star formation and black hole accretion at high redshift 
might differ from their low redshift counterparts, detailed calculations
and simulations have not yet been done.  We use low redshift calculations
and measurements as a guide.  We stress that we are most interested in
the effects of the time dependence, as the unknowns of
the detailed modeling are vast.

A major merger to a halo of mass $m$ will add a total number of photons
$N_{\gamma}(m)$ over a time $t_{ion}$.  These
$N_\gamma$ photons can ionize the hydrogen in a region with mass $M$
obeying 
$N_{\gamma} = \frac{M}{m_p}
\frac{\Omega_b X}{\Omega_m}$ where $X = 0.76$ and
$m_p$ is the proton mass.\footnote{In practice there will be a small correction,
perhaps of order 10\% due to the possibility that some of the photons will
ionize Helium instead, which should be kept in mind.  It can be thought of
as a rescaling of our fiducial input parameters.  We thank the referee
for pointing this out.}  
This is the amount of mass $M$ that can be ionized over the whole time
$T$, giving, if $M = \zeta m$, 
\eq
\label{zetagamma}
\zeta = \frac{N_{\gamma}}{m} \frac{m_p}{X} (\frac{\Omega_b}{\Omega_m})^{-1}
\end{equation}
We will want to compare the contributions to photons at some given
time, thus we will look at all the photons being contributed assuming
a constant\footnote{
More accurately,
the accretion onto a black hole, if Eddington as is commonly assumed,
tends to increase with time until all the infalling material is gone, while the 
starbursts tend to decay with time until all the gas is gone.  However other
estimates of black hole accretion have the accretion also slow down
with time as the gas supply is decreased.} 
production rate, i.e. $\zeta_t = \frac{\zeta}{t_{ion}}$.  The
times $t_{ion}$  for black hole accretion and starbursts differ in general.

\subsection{Star Formation and Starbursts}

Photon production due to star formation can be
estimated via (e.g. Loeb, Barkana \& Hernquist \cite{LoeBarHer05})
\eq
\label{ngstars}
\frac{d N_\gamma,stars}{dt} = 
\frac{m}{m_p} \frac{\Omega_b}{\Omega_m} \frac{N_{ion}}{t_*} \; .
\end{equation}
leading to
\eq
\label{zetngstars}
\zeta_t(m_i,m_0,t_i,m,t) =
\frac{1}{X} \frac{N_{ion}}{t_*} \;  (stars)
\end{equation}
Here $N_{ion}$ is the overall number of ionizing photons per baryon in
the halo.
We assume that the starbursts change the rate from a constant
quiescent rate to a constant starburst rate for a 
time $t_* + t_{burst}$, until the stars due to the
starburst are gone.\footnote{Note
 that the photon production rate estimate in FO05
and in FMH05
has $\zeta$ time independent, and assumes that all mass
going above $10^4K$ is instantaneously converted into photons,
which is another common approximation, e.g. Somerville \& Livio 
\cite{SomLiv03}.}

For $N_{ion}$ our parameters are similar to Loeb et al \cite{LoeBarHer05}. 
We take a star lifetime $t_* = 80 Myr$.  Showing our choices in parentheses\footnote{We choose $f_{esc} = 0.05$ cf.
Wyithe \& Loeb \cite{WyiLoe03}), canonical choices vary by a factor of 10. 
Loeb et al use $N_{\gamma,bary} = 4300$ for a Scalo IMF of metallicity
1/20 of solar.
},
the efficiency with which baryons are incorporated into stars is $f_*$(=0.1),
and each baryon in a star produces $N_{\gamma,bary}$(=4400) photons, 
with $f_{esc}$(=0.05) of them escaping.  Combining these we get
\eq
\begin{array}{lll}
N_{ion} &= N_{\gamma,bary} f_* f_{esc}  & m>m_{ref}(z) \\
        &= N_{\gamma,bary} f_* f_{esc} (\frac{m}{m_{ref}(z)})^\alpha  
& m<m_{ref}(z) \\
\end{array}
\end{equation}
The power $\alpha = 2/3$ is to take into account the scaling of star formation
found in low stellar mass 
galaxies (below $m_{stellar} = 3 h \times 10^{10} h^{-1} M_\odot$) 
by Kauffmann et al \cite{Kau03}, we assume that it is the same
for both starbursts and quiescent star formation as in the $z=0$
sample low mass halos are not merging very often.  We will call this scale
$m_{ref}$ in the following, our choices for it are discussed below.
A similar mass dependence was introduced into $\zeta(m)$ by FMH05, but
without the transition mass $m_{ref}$.

With a starburst, $f_*$ will
be larger (more gas will go into stars) for a time $\sim t_{burst}+t_*$
(that is, after the merger the number of baryons going into stars will
increase, and these stars will add extra photons for their entire lifetime).
The average $f_{*,burst}$ at this time is $f_{*,burst} t_*/(t_{burst} + t_*)$.
Simulations of mergers at low redshifts provide some suggestions for $f_*$
for a starburst, e.g. a range of 65\% - 85\% of the total gas (
Mihos \& Hernquist \cite{MihHer94,MihHer96}) or $\geq$80\% of the
cool gas (Springel, Di Matteo and Hernquist \cite{SprDiMHer05})
going into stars (see also the studies of gas available at early times
by Machacek, Bryan \& Abel \cite{MacBryAbe03}).  
Observationally, starburst star formation is fit by a decaying
exponential with time scales as short at 20 Myrs in some examples
but lasting up to hundreds of Myrs (e.g.Papovich, Dickinson \& 
Ferguson \cite{PapDicFer01}, Shapley et al \cite{Sha01}, Conselice 
\cite{Con05}) with similar time ranges used to analyze observed
samples (Kauffmann et al \cite{Kau03}). 
 
We only consider Pop II stars, although it would be
straightforward to generalize beyond this in our formalism.
The transition redshift between creating Pop III stars and creating
Pop II stars has been estimated to be $z=15$
(Yoshida, Bromm \& Hernquist \cite{Yos04}, see Fang \& Cen
\cite{FanCen04} for more discussion of constraints on this
transition); however a wide range of final redshifts
where Pop III stars might be important for reionization have
been suggested (between $z=10$ and $z=20$
e.g. Wyithe \& Padmanabhan \cite{WyiPad05},
Wyithe \& Cen \cite{WyiCen06}).
(There are even searches for evidence of
Pop III stars at redshifts down to $z \sim 6$
(Scannapieco et al \cite{Sca05}).)
As a result we do not consider redshifts
before $z \sim 12$ (which are active, for our choices of
time scales, from mergers at $z \sim 15$).

There are expected differences at high redshift, for example the
higher densities $\sim (1+z)^3$ lead to much shorter cooling times,
as well as lower metallicities, thus these guesses are just that,
however they are a starting point for estimating the size of
time dependence of $\zeta$.

\subsection{Black Hole Accretion}

We now turn to photon production due to black hole accretion of mass
 $\Delta m$.
We take (again assuming
a constant photon production rate and using the notation of
Salvaterra, Ferrara \& Haardt \cite{SalHaaFer05}), 
\eq
\begin{array}{ll}
\frac{dN_{\gamma,bh}}{dt} &= 
\Delta m  c^2 \frac{f_{UV}}{\langle h \nu \rangle} \frac{\epsilon}{t_{acc}}
\end{array}
\end{equation}
which leads to (using the time dependent version of eqn. \ref{zetagamma})
\eq
\zeta_t(m_i,m_0,t_i,m,t) = 6.9 \times 10^7
\frac{\Delta m}{m X} \frac{\Omega_m}{\Omega_b} \frac{f_{UV}}
{\frac{\langle h \nu \rangle}{13.6 eV}} 
\frac{\epsilon}{t_{acc}}
\end{equation}

The black hole photon production formula only applies
to halos hosting a black hole (i.e. with mass above $m_{bh,min}(z)$) and
includes the effects from UV photons only. 
The factor $\frac{f_{UV}}{\langle h \nu \rangle} = 0.1-0.2 \; {\rm ryd}^{-1}$ 
(e.g. Madau et al \cite{Mad04}).  We are assuming
that $f_{UV}$ takes into account any local absorption, i.e.
it is the flux from the whole region due to the mass accretion.
The efficiency is usually taken to be $\epsilon = 0.1$.
The mass accreted $\Delta m$ is often specified in terms of halo mass
in the range $10^{-3} - 10^{-5} m$.  Assuming the mass accretes at the
Eddington rate 
\eq
\dot{m_{bh}} = m_{bh}/t_S \; , \; \; t_{S}= 45  Myr (\epsilon/0.1)^{-1} \; ,
\end{equation}
and commonly used black hole mass to halo mass relations, $m_{bh} \sim
10^{-3}-10^{-4} m$, this
gives lifetimes $t_{acc}$ from $0.01-10$ $t_S$. We will take a relatively
short accretion time, 0.1 $t_S$, as our fiducial time 
$t_{acc}$ and $m_{bh} \sim 10^{-4} m$.\footnote{ Many different concerns suggest a
shorter $t_{acc}$.  Relating the black hole
mass instead to the halo velocity,
$m_{bh} \sim 10^{-4} (1+z)^{3/2} m$ i.e. including 
redshift dependence (Bromley, Somerville \& 
Fabian \cite{BroSomFab04},Wyithe \& Padmanabhan \cite{WyiPad05}),
lowers $t_{acc}$.
Accretion at super-Eddington rates (argued necessary to get
super massive black holes by redshift 6 or 7, 
e.g. Haiman \cite{Hai04}) also gives a shorter $t_{acc}$.  In addition,
shorter accretion times are also expected from some 
low redshift constraints based on the luminosity function
(e.g. Wyithe \& Padmanabhan \cite{WyiPad05}). }

For our purposes we consider all halos above some minimum
mass $m_{bh,min}(z)$ 
to host black holes, and all halos below it not to host black holes,
and ignore the scatter in this relation.  We tried to estimate this
minimum mass by using methods already found in the literature.
We have two choices corresponding seed black holes at $z=24$  in
halos which are $3.5 \sigma$ and $3 \sigma$ fluctuations, following
Madau et al's approach \cite{Mad04}, see the appendix for more discussion
of this choice and constraints. 
The two cases used for minimum halo masses hosting
black holes and $m(10^4 K)$ are shown as a function of redshift in Fig. 1.
Our fiducial model takes black hole hosts as 
$3.5\sigma$ fluctuations at $z=24$, concordance cosmology, 
and the parameters in eq. \ref{eq:params}.
\begin{figure*}[htb]
\begin{center}
\resizebox{5.5in}{!}{\includegraphics{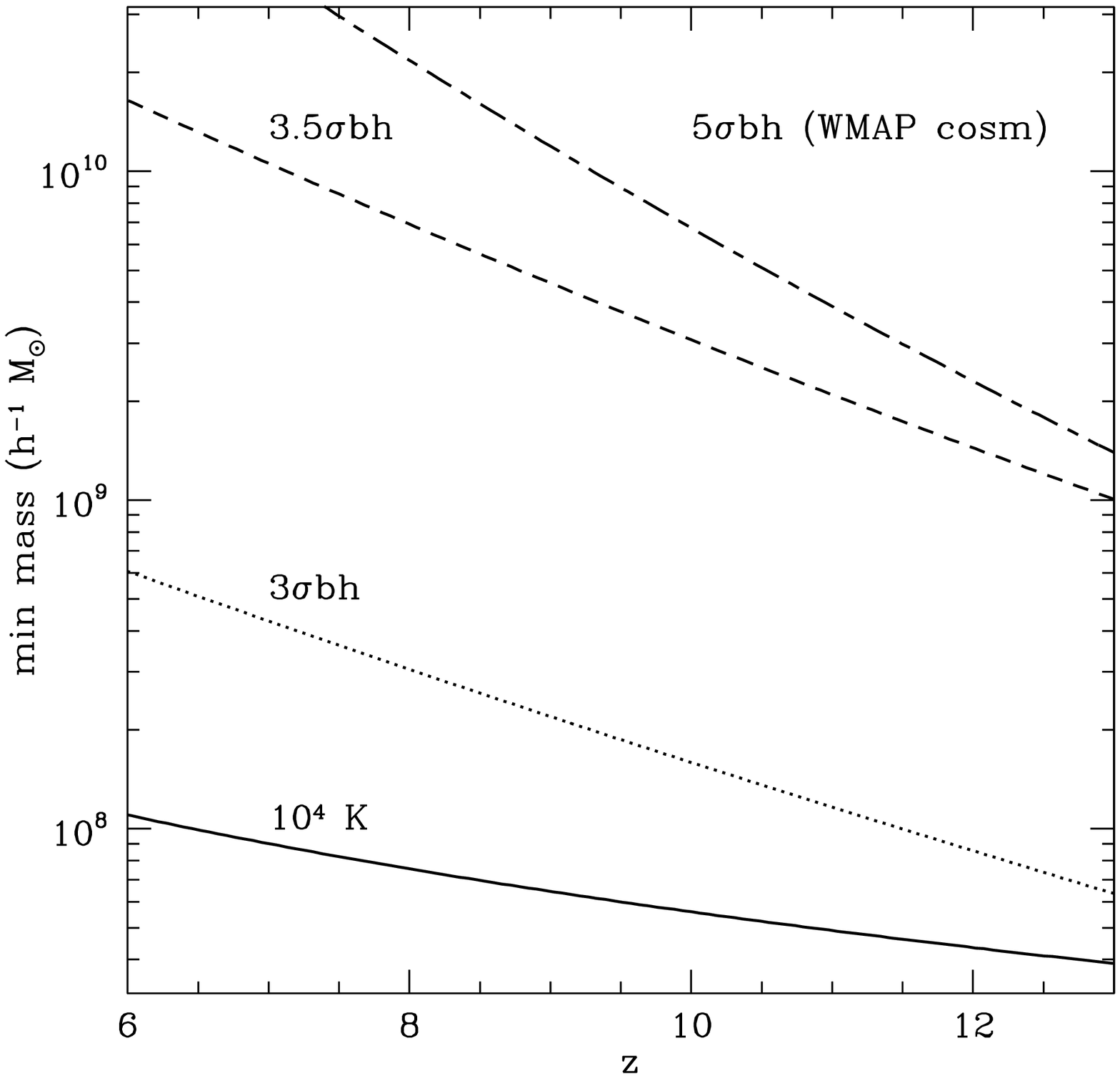}}
\end{center}
\caption{Minimum halo masses as a function of redshift.  The lowest 
solid line
is the mass of a $10^4 K$ halo. The top 3 lines
show three models for the minimum mass for
a halo to contain a black hole.
They come from choosing black hole hosting halos at $z=24$ which are
respectively $3 \sigma$ (dotted), 
$3.5 \sigma$ (dashed),
or $5 \sigma$ (dot-dashed, uses WMAP cosmological parameters) peaks;
see the appendix for details.
Our fiducial model uses the $3.5 \sigma$
seeds at $z=24$.}
\label{fig:minmass}
\end{figure*}
As mentioned earlier, we also explored some models with WMAP
cosmological parameters: here
a detailed application of the black hole
halo constraints (some of which are described in the appendix) 
has yet to be done.  However, in this case
 $3 \sigma $ and $3.5 \sigma$ halos at $z=24$ 
have masses $11.3 h^{-1} M_\odot$ and $600 h^{-1} M_\odot$
respectively, which is difficult to reconcile (especially in the
former case) with initial black hole masses
of $\sim 100 h^{-1} M_\odot$.   
We took one exploratory example:
a black hole in 5 $\sigma$ 
peaks at $z = 24$ (halos of mass $\sim 6 \times 10^5 h^{-1} M_\odot$).  
The motivation for the choice was that at least initially 
$m_{bh} \sim 10^{-3}-10^{-4} m_{halo}$ and this limit is also shown in 
Fig. \ref{fig:minmass}.

\subsection{Recipe for Ionizing Photons}\label{sec:recipe}

Combining the above,
we have the following recipe for $\zeta_t(m_i,m_0,t_i,m,t)$ due to a
$m_i$ halo merging to mass $m_0$ at time $t_i$ which
later ends up in $m$ at time $t< t_i + t_{ion}$:
\begin{table}[hbtp]
\centering
\caption{``Recipe'' for $\zt$}
\begin{tabular}
{l|c|c|c|} 
\\
{\rm cause} & {\rm major merger} & $m  \zeta_t$ & $t_{ion}$ \\ \hline
{\rm quiescent} & -- &
$\beta_{*,q} \frac{m^{1+\alpha}}{ m_{ref}^\alpha}$  & $t_*$
= 80 Myr \\ \hline
{\rm starburst} & $>1:3$ &
$(\beta_{*,sb}-\beta_{*,q}) \frac{m_0^{1+\alpha}}{m_{ref}^\alpha} $ & $t_{burst} +t_*$
= 100 Myr \\ \hline
{\rm black holes} & $> 1:10$ &
$\beta_{bh} m_0 $ & $t_{acc}= 0.1 \, t_S = 4.5$ Myr \\ \hline
\end{tabular}
\end{table}
Besides the merger being within the $t_{ion}$
of interest (starburst or accretion), the $\zt$ are also zero 
unless the masses also satisfy certain conditions.
For the mass range cutoff for a major merger we will use 
1:3 for starbursts (many ranges are used in the
literature)  and 1:10 for black hole accretion
(e.g. Madau et al \cite{Mad04}).   For
black holes $m_i > m_{bh,min}(z(t_i))$, for quiescent and starburst star
formation $m_0 > {\rm max}(m(10^4K(t_i)),m_{0,min}(t_i))$.

Then $\beta_*, \beta_{bh}$ encapsulate all the non-mass dependent factors:
\eq
\displaystyle
\label{eq:params}
\begin{array}{ll}
\beta_{*,q} &= f_* N_{\gamma,bary} \frac{f_{esc}}{t_*} 
\frac{1}{X} \\
 &= \frac{.36}{\rm Myr} \frac{f_*}{0.1}
\frac{N_{\gamma,bary}}{4400} \frac{f_{esc}}{0.05} \frac{80 {\rm Myr}}{t_*} \\
\beta_{*,sb} 
&= \frac{1.5}{Myr} \frac{f_{*,burst}}{0.5} \frac{N_{\gamma,bary}}{4400}
\frac{f_{esc}}{0.05}
\frac{\frac{t_*}{t_{burst} + t_*}}{0.8} \\
\beta_{bh} &= 6.9 \times 10^7 \frac{\Delta m_0}{t_{acc}m_0} \epsilon 
\frac{f_{UV}}{\frac{\langle h \nu \rangle}{13.6 eV} }
\frac{\Omega_m}{\Omega_b X}\\
&=
\frac{15.3}{Myr} \frac{f_{UV}/\langle h \nu \rangle}{0.1 ryd^{-1}}
(\frac{m_{bh}/m}{10^{-4}}) (\frac{\epsilon}{0.1})
\end{array}
\end{equation}

Our fiducial parameters are shown in the second line of each equation.
Our fiducial $\alpha = 2/3$.
The reference mass $m_{ref}$ 
is taken to change with redshift,
\eq
m_{ref}(z) = 4.3 \times 10^{10} [\frac{\Omega_m(0)}{\Omega_m(z)}
\frac{\Delta_{crit}(z)}{18 \pi^2}]^{-1/2} (\frac{1+z}{10})^{-3/2}
h^{-1} M_\odot \; ,
\end{equation}
 (e.g. Wyithe
\& Loeb \cite{WyiLoe03}\footnote{They interpret the $m^{2/3}$ behavior
as due to a potential well
effect.  We
have taken their default parameter values for $v_* = 176$ km s${}^{-1}$
and $\Delta_{crit} = 18 \pi^2 + 82 x -39 x^2$,
$x = \Omega_m(z) -1$, $\Omega_m(z) = \Omega_m(0) (1+z)^3 /[\Omega_m(0)
(1+z)^3 + \Omega_\Lambda]$.  At
$z\geq 9$, $\Delta_{crit}/18 \pi^2  \sim \Omega_m(z) \sim 1$.}).
Some numerical studies have found a transition mass $m_{ref}$
which doesn't change
significantly with redshift (Keres et al \cite{Ker05},
the corresponding 
transition mass is also almost a factor of 10 smaller at $z=0$). 

For $\beta_{bh}$ we estimate $\frac{\Delta m_0}{m_0 t_{acc}}$
assuming an Eddington rate and $t_{acc}/t_{S} \ll 1$ so
that $\Delta m = m_{bh} (e^{t_{acc}/t_S} -1)\sim m_{bh} t_{acc}/t_S$ and we take
the
black hole mass proportional
to its host halo mass, $ m_{bh} = 10^{-4} m_0$. 
The suggested relation of black hole mass to halo velocity
mentioned earlier
would give a larger ratio of black hole mass to halo mass and a bigger
effect from black holes.
In addition, super Eddington accretion will also shorten
$t_{acc}$ and increase $\beta_{bh}$.
Note that changing 
$t_{*}+t_{burst}, t_{acc}$ also changes the range of integration for
$t_i$ in calculating $\zeta_t$,
not only the prefactors $\beta_*, \beta_{bh}$.  
For starburst star formation $f_{*,burst}$ gives an effective $f_*$ of 
$f_{*,burst} t_*/(t_* + t_{burst})$. 

With these assumptions, the $\zt$ in
equation \ref{photonrate}
are a set of constants for any given $m_0$ (or $m$ for quiescent star
formation). 
They are only nonzero when $t_i, m_i, m_0$ are in
the right range, and for starbursts, black hole accretion or
their combination.   These constants 
were denoted as $\zeta_t^a(m_0)$ in section \ref{scatter}.
All of these numbers rely on a huge number of
estimates of unknowns, i.e. the contributions of
starbursts, black hole accretion and their relative strengths.  As mentioned
earlier,
ideally one could turn this around and use it to estimate
the contributions from starbursts and quasars, but many uncertainties
are involved.  Here we are
interested in the sporadic and time dependent 
nature of the mergers changing photon
production rates.

\begin{figure*}[htb]
\begin{center}
\resizebox{5.5in}{!}{\includegraphics{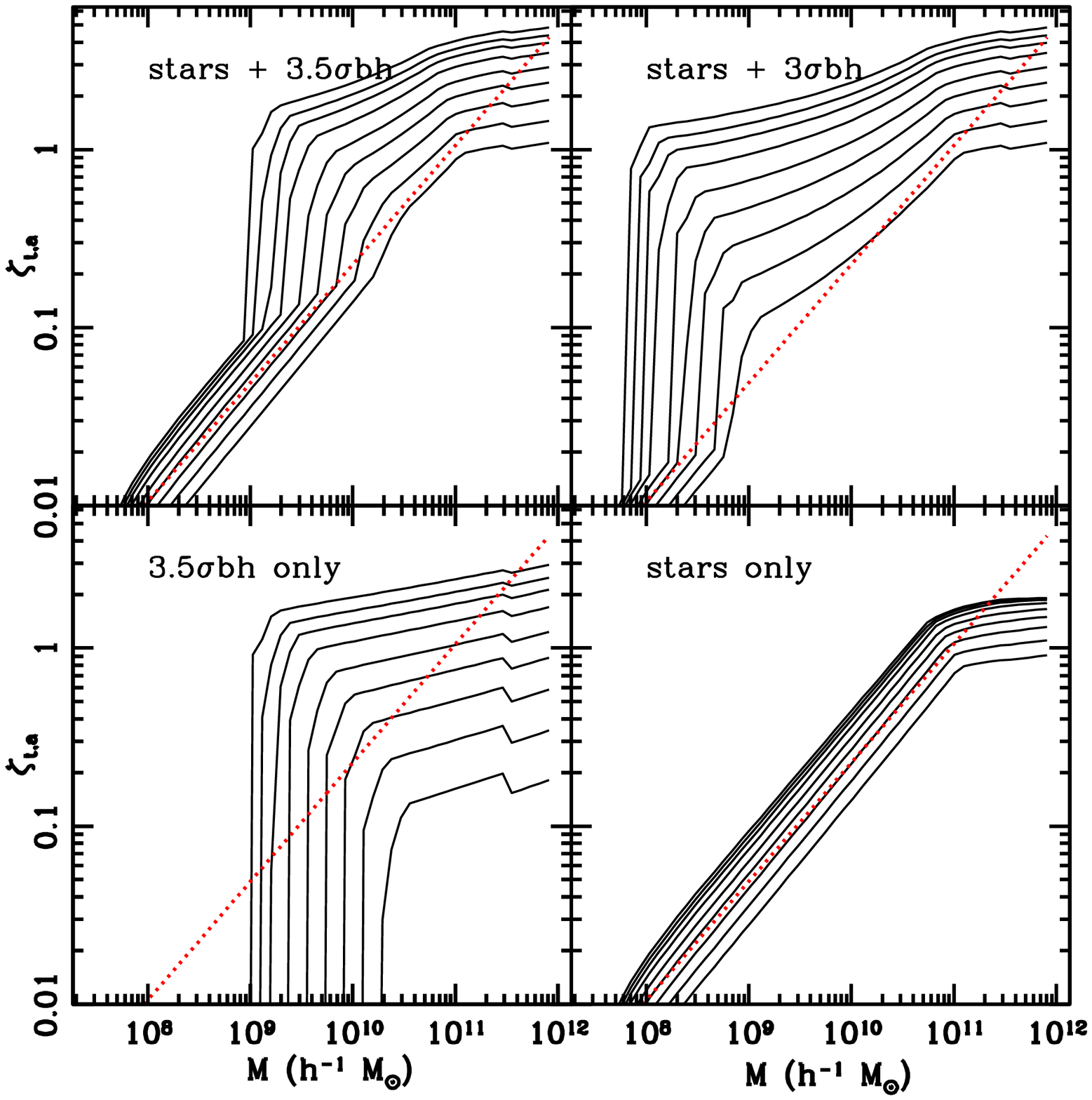}}
\end{center}
\caption{Redshift and mass dependence of $\zeta_{t,a}(m)$, lines are
$z = 5.9,7.1,8.0,9.0,10.2,11.2,11.9,12.6.13.4$, bottom to top
(mergers are more common in the past, increasing photon production and thus
$\zeta_{t,a}(m)$).  Top left: stars
and 3.5 $\sigma$ black holes, top right: stars and $3 \sigma$ black holes.
Bottom left: 3.5 $\sigma$ black holes only. Bottom right, stars only.
The dotted line in each case is the slope that would
arise from a time independent broken power law as described in eq. \ref{zfmh}.
Mergers steepen the slope slightly ($ m^{2/3} \rightarrow \sim m^{0.7}$)
at every redshift.}
\label{fig:fourzeta}
\end{figure*}

\begin{figure*}[htb]
\begin{center}
\resizebox{5.5in}{!}{\includegraphics{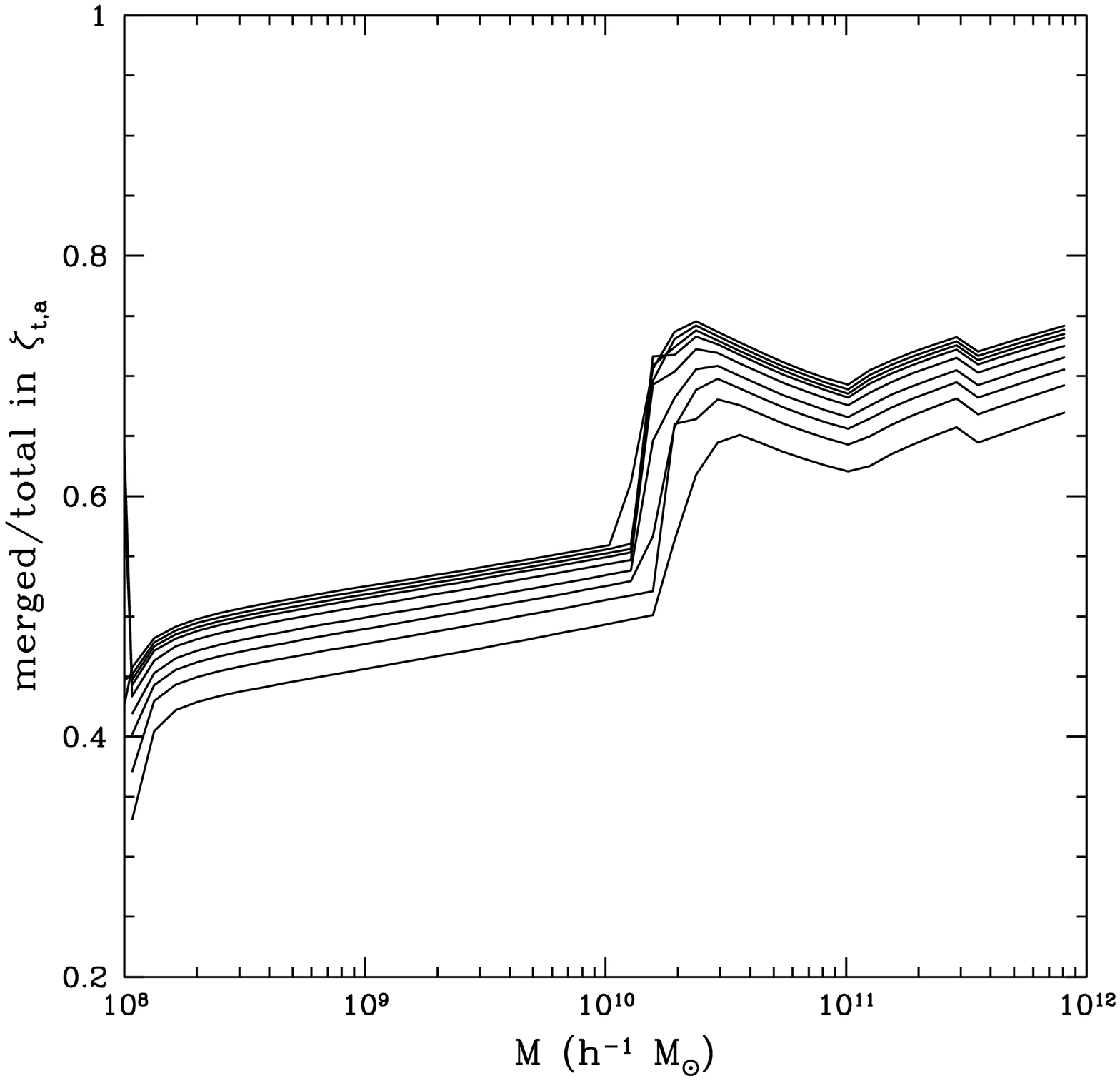}}
\end{center}
\caption{Ratio of photon production rates:
merger-produced/(merger produced and
quiescent) as a function of mass, for the same redshifts
as Fig. \ref{fig:fourzeta}, low to high redshift going
bottom to top. The small wiggle
immediately above the sharp rise is a numerical artifact.}
\label{fig:mercont}
\end{figure*}

Using these definitions, $\zeta_{t,a}(m)$ (equation \ref{photonrate})
is shown in Fig. \ref{fig:fourzeta} for two black hole
assumptions for a series of different times. 
The black hole and star formation contributions are shown separately as well.
The quiescent contributions to $\zeta_{t,a}$ roughly scale with the dotted
line in each case, but do not change with redshift.
The ratio of merger-induced photon contributions to total photon
contributions for our fiducial model  and 
for the same redshifts is shown in Fig. \ref{fig:mercont}.
The source mergers have a similar photon production
rate to the quiescent rate,
increasing with increasing mass and redshift. There
is also a sharp increase in the merger contribution relative to
the quiescent contribution once the masses are large enough to host
black holes.  
Although the black holes have a very large contribution per halo for
high mass halos, the sharp decline in numbers of halos as mass increases
somewhat limits their effects.

\section{Consequences for bubbles}
The framework and prescriptions of the previous two sections can
now be combined to predict the resulting properties of the ionized bubbles.
In practice there are two steps: $\zt$ is used to find
the required overdensities for
given bubble masses $M$, i.e. $\delta_M$.
The barriers $\delta_M$ then determine the bubble size distributions
and the time
evolution of the ionization fraction.

Our fiducial model, shown unless otherwise stated, is
for the $3.5 \sigma$ seed black holes at z=24
as described earlier and in the appendix, 
with parameters given in eqn. \ref{eq:params}.
We also considered
$3\sigma$ seed black holes at $z=24$, $\alpha = 0$,  $m_{ref}$ constant,
$f_* f_{esc}\rightarrow 4f_* f_{esc}$, no stars,
no black holes.  As mentioned above, we also varied the cosmology in
several ways.  We considered $\Omega_b h^2 = 0.0225$ leaving
all else fixed and cosmological tilt $n= 1.05$ leaving all else fixed.  
The latter model
seemed to have very similar effects to the one increasing $f_* f_{esc}$.
We took some combinations of the variations with each other as well. 
 The two lower integration
limits for $m_0$, ($m_{0.min} = 0.1m$ or $m_{0,min} = m(T=10^4K)$) 
gave very similar results when $\alpha = 2/3$.  
We considered minihalo and MHR recombination for all the models,
and MIPS recombination for the fiducial model and 5 representative variations.
One other quantity we must fix is an initial condition; this is discussed
below.

\subsection{Barriers}

To find the required $\delta_M$ given a bubble mass $M$ we use 
the generalization of equation \ref{zetaf}, equation \ref{fullzetr}:
\eq
\int^t dt' [(\zeta f)_{t'} - A_u (1+\delta_{R,M}(t'))
\frac{M_{ion}}{M}C(R_{ion}(t'))]  = 1 \; .
\end{equation}
Because time dependence is built in, we either need to integrate all
the way back to the time of the very first ionizing photon production or 
include some initial condition of global ionization fraction.  We choose
to fix an initial condition at $z=12$.  At $z=12$ halos are only
active from mergers since $z \sim 15$ with our choices of relaxation times.
As discussed earlier $z = 15$ is the earliest redshift Pop II star
dominance (and thus our assumptions) might be expected to hold.

The distribution corresponding to the initial photons is taken to be
a time independent $\zeta(m)$ similar to the form given by FMH04 \& FMH05,
so either $\zeta_{FMH}$ is constant or
\eq
\label{zfmh}
\zeta_{FMH} \propto \{ 
\begin{array}{ll}
(\frac{m}{m_{ref}})^{2/3} & m < m_{ref} \\
const & m > m_{ref} 
\end{array}
\end{equation}
in accordance with the mass dependence of our $\zeta_t$ ($\alpha = 0,2/3$).
Then $\zeta_{FMH}$ is normalized by setting the global ionization fraction
\eq
\lim_{M\to \infty,\delta_M \to 0}
\int dm \zeta_{FMH}(m) P_1(m,t_{bc}|M,\delta_M) = \bar{x}_{i,bc} \; .
\end{equation}
At a later time, the ions present in a region of mass $M$ with
overdensity $\delta_M$ are a combination of the ions produced in
that region at the initial time plus those produced since, minus recombinations:
\eq
\begin{array}{l}
\int dm \zeta_{FMH}(m,t_{bc}) P_1(m,t_{bc}|M,\delta_M)  \\
\; \; \; \; +
\int_{t_{bc}}^t dt' [(\zeta f)_{t'} - A_u (1+\delta_{R,M}(t'))
\frac{M_{ion}}{M}C(R_{ion}(t'))]  = 1 \; .
\label{eq:barrier}
\end{array}
\end{equation}
For our fiducial model, we start with an ionization
fraction $\bar{x}_i(z=12)$ of almost zero ($10^{-3}$).  This should isolate
the evolution of bubble properties due to the change in photon
production since the initial time.  We also experimented with some
non-negligible initial photon distributions at $z=12$ 
($\bar{x}_i = 10^{-2},10^{-1},0.5, 0.14$) and an initial condition
of $\bar{x}_i = 10^{-3}$ at $z = 16$ when halos are active
from mergers since $z=23$.  This last corresponded to
$\bar{x}_i = 0.16$ at $z=12$.  In practice at $z=16$ Pop II stars are
much less likely and our assumptions about photon production rates
suspect, as a result this model was not our fiducial model, but is 
useful for studying effects of initial conditions.

To find $R_{ion}(t)$ the integral is divided into 700 steps from the initial
time to the redshift of interest.  At each time, the instantaneous
change in ionized mass fraction $(\zeta f)_t dt$ is added to
the mass fraction already present (the integral up to that
time) to find $M_{ion}/M$, the ionized mass fraction.
In the case where more photons were produced
than hydrogen atoms in the region, we took $M_{ion}/M \equiv 1$
(allowing it greater than one had a neglible effect). We took
$R_{ion} = [\frac{M_{ion}}{\frac{4 \pi}{3} \bar{\rho}(1+\delta_{R,M}(t))}]^{1/3}$,
however replacing $\frac{3M_{ion}}{4 \pi} \rightarrow 
M_{ion}$ or even $\frac{3M}{4 \pi}$ had very small
effects on the resulting barriers we found;
the biggest effect is due to the overall coefficient for 
recombinations,
$\frac{M_{ion}}{M}$.  Larger numbers of steps gave indistinguishable results.

\begin{figure*}[htb]
\begin{center}
\resizebox{5.5in}{!}{\includegraphics{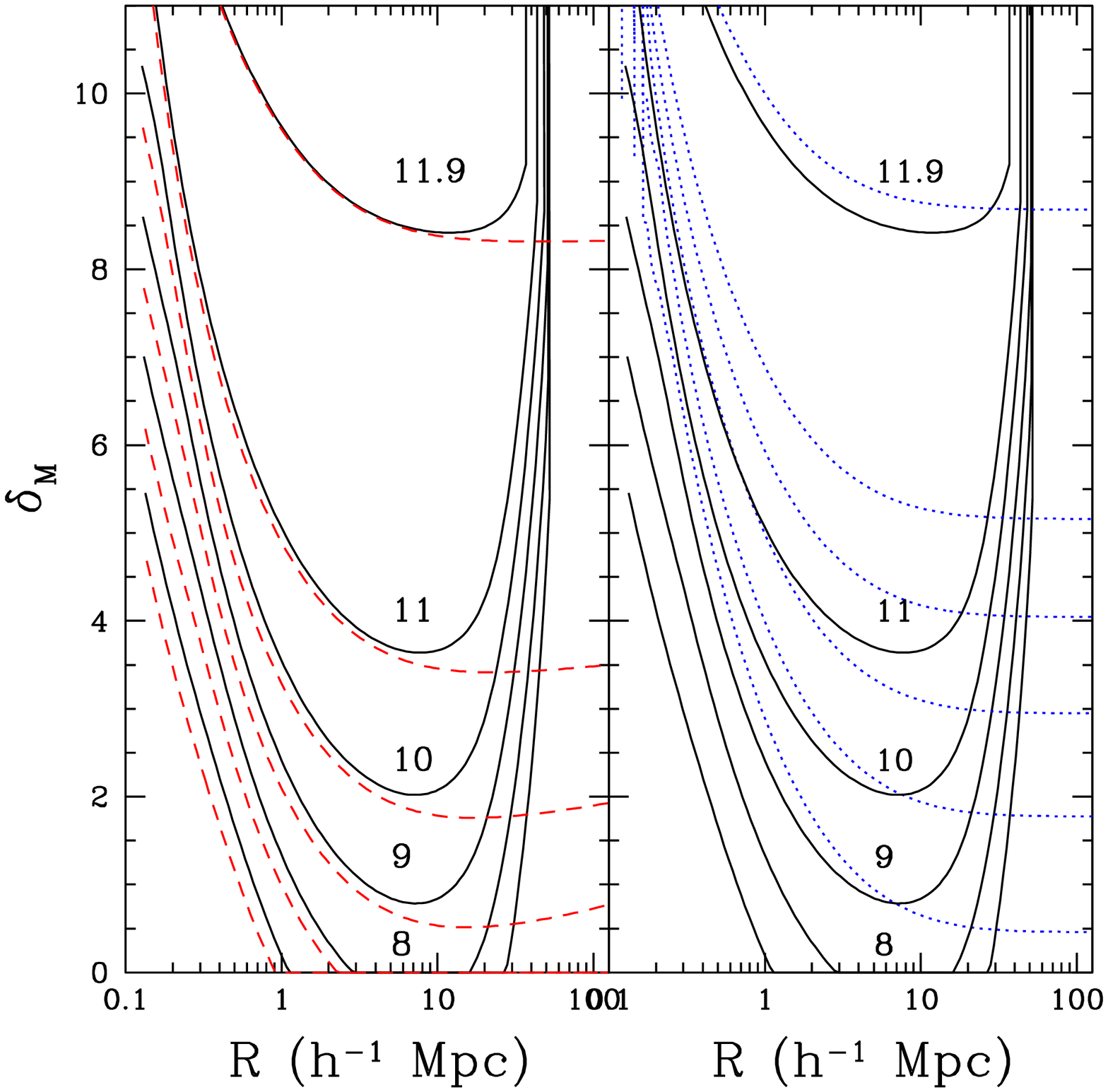}}
\end{center}
\caption{The barriers $\delta_M$ found at $z = 7, 8, 9, 10, 11, 11.9$ for 
our fidicial model.
The solid lines (left and right)
 use minihalo recombination (eqn. \ref{eq:mini} in text),
the dashes (left) use MHR recombination (eqn. \ref{eq:cmhr}),
and the dots (right) use MIPS recombination (eqn. \ref{eq:mips}).  The
redshifts for minihalo recombination are labelled, for MHR or MIPS they
coincide at high $z$ and then can be deduced by counting lines down.}
\label{fig:barriers}
\end{figure*}

We show in Fig. \ref{fig:barriers} 
examples of the resulting barriers $\delta_M$ 
at $z$=7, 8, 9, 10, 11 and 11.9 for our fiducial model and 
the three recombination prescriptions.
At large radius, $C_{MHR}(R)$ tends to approach a
constant, $C_{MIPS}$ is 
constant, and $C_{mh}(R)$ increases exponentially.
Thus the barriers for minihalo recombination are cut off at large
radius.  MIPS and MHR recombinations allow bubbles to
get very large, if there is sufficient photon production, to the point
where consistency questions
involving e.g. travel times within the corresponding bubbles arise. 
For plots of many of the models we thus choose to only include the cases with
radii $< 100 h^{-1} Mpc$, models with much larger radii do appear but
are difficult to interpret. 

Our time dependent approach produces results
which interpolate between the two limits used in earlier works (e.g.
FZH04,FO05,FMH05): they included
implicit recombinations at early times, and imposed equilibrium between
instantaneous photon production and recombination rates at late times.
These two limits, in terms of $\delta_M$, give the following:
the early time approximation produces a barrier
similar to ours for small $M$, approaching a
constant (horizontal line as function of $M$) for large $M$.
The late time (larger $M$) limit gives a barrier $\delta_M$ 
that is very close to a vertical line (see for
example FO05, Fig.7), combining the early and late time limits gives
the barrier we found which rises at both low and high $M$.
Physically, for small bubbles, the photon mean free path
exceeds the bubble radius, thus photons escape freely and
recombinations do not play an important role; this is expected to
happen at early times.  As bubbles grow in size, the barriers for
ionized regions decrease as a function of bubble radii, since more and
more mass (and therefore ionizing photons) can be included in the
region. At late times, bubble radii grow to exceed the photon mean
free paths and recombinations limit the growth of bubbles. 
Of the three assumed IGM gas density distributions, the minihalo
and MIPS models put more stringent constraints on the bubble growth 
than the MHR model.  The stronger effects of minihalo recombination
compared to MHR recombination were 
also observed in the models equating instantaneous recombination and
photon production (FMH05 and FO05).

\subsection{Bubble Size Distribution}

The barriers found as a function of bubble radii translate
directly into a bubble ``mass function'' $n_b(M,\delta_M)$ which gives
the number density of bubbles with masses between $M$ and $M + dM$.
Descriptions of how to derive
the resulting bubble mass function $n_b(M,\delta_M)$ for
linear barriers can be found e.g. in 
Sheth \& Tormen (\cite{She98}) and McQuinn et al (\cite{McQ05}).
As the barriers here are non-linear in $\sigma^2(M)$, we find
$n_b(M,\delta_M)$ by simulating 4000 random walks directly for
each $(M,\delta_M)$ combination. For linear barriers, 4000 steps
reproduced the ionization fractions to a 5-10\% percent once 
$\bar{x}_i > 0.01$.  

The quantity $n_b(M,\delta_M)dM = M n_b(M,\delta_M) d \ln M$ 
counts the number of bubbles with log mass $\ln M$.  One might want this
quantity if one knows one has a collection of bubbles and wants to know 
the size distributions of these bubbles.  However, for many questions it is
of interest to know the fraction of mass in the universe in ionized bubbles
with $\ln M$ in the range $\ln M$ to $\ln M + d \ln M$, i.e.
$\frac{M^2}{\bar{\rho}} n_b(M,\delta_M) d \ln M$.  This peaks at a larger
mass than the former quantity due to the added factor of $M$.
The peak of this distribution corresponds to a mass $M_c$ with
a characteristic radius $R_c = [3 M_c/(4 \pi \bar{\rho} 
(1+\delta_{R,M_c}(t)))]^{1/3}$.  This radius (used also in earlier work,
e.g. FMH05) denotes to the bubble size which contains the
largest fraction of the ionized mass.    The quantity
$\frac{M^2}{\bar{\rho}} n_b(M,\delta_M) d \ln M$ is dimensionless and can
also be thought of
as probability distribution for $P(\ln M) d \ln M =
P(\ln R) d \ln R$ (however it integrates to $\bar{x}_i$ rather than to one).
This distribution and $R_c$ are
the quantities shown/derived in previous work, e.g. FMH05.  The
one big difference is that they
divide by the overall ionization fraction $\bar{x}_i$.  I.e. our plots give
the fraction of the {\it total} mass which is
in the ionized bubbles with the given $\ln M$.  Our motivation for this
is that the Monte Carlo calculations rely on counting paths crossing
the barriers $\delta_M$--with more fractional scatter in counts when
there are fewer counts.  By not including their factor
$1/\bar{x}_i$ one
can read off immediately which curves have the largest such sampling
errors (i.e. the ones with the lowest heights).

\begin{figure*}[htb]
\begin{center}
\resizebox{5.5in}{!}{\includegraphics{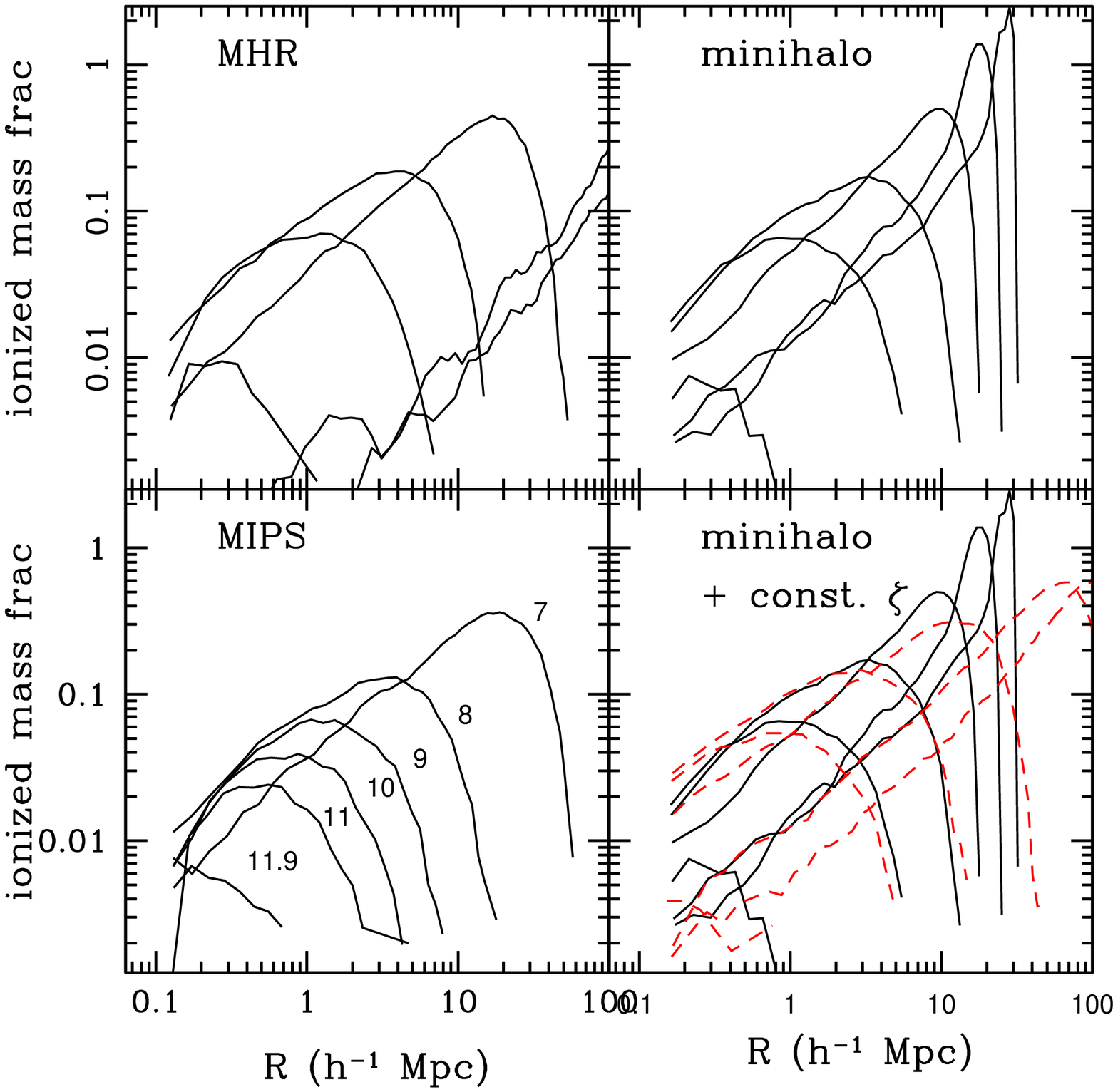}}
\end{center}
\caption{Top left and right and bottom left: ionized 
mass fraction ( $3 \frac{M^2}{\bar{\rho}} n_b(M,\delta_M)$,
i.e. fraction of total mass), in bubbles of radius $R$
for three recombination prescriptions.
The lines are for different redshifts (11.9, 11, 10, 9, 8, 7), 
the smallest height curves are at the earliest time.
The MHR recombinations are weakest and thus result in the largest bubbles,
extending at late times to sizes where many other concerns arise as
to interpretation of the model.  At bottom right, the minihalo recombination
model is shown again (solid line) along with a time independent zeta 
$\sim m^{2/3}$ model (eq. \ref{zfmh}) having
the same $\bar{x}_i$ at each redshift (dashed line).
The time independent models 
have comparable central values of $R$ for high redshifts and low ionization
fractions, i.e. at early times when recombinations aren't very
important. Generally their radial
distributions are wider as their recombinations are only implicit.}
\label{fig:radii}
\end{figure*}

In Fig. \ref{fig:radii}, 
we show $R$ vs. $3 \frac{M^2}{\bar{\rho}} n_b(M,\delta_M)= P(\ln R(M))$ for the 
sequence of different redshift barriers of Fig. \ref{fig:barriers}
above.
The trends shown here were seen in our other models as well.
For all the models, the radii and ionized mass fraction grow with time.
The late time shape of the distribution in $\ln R$ has a strong dependence upon
the assumed form of recombinations.  This is also shown in Fig. \ref{fig:varyic}
below, lower left.

For minihalo recombinations the
width of the probability distribution for $R$ narrows as $R$ increases
and the universe ages.  This schematically shows the progress of
reionization: at early times bubbles are small and have a large range
in sizes, depending more strongly on the local density and ionizing
photon production where the bubbles live.  As the global ionization
fraction grows, or as the bubble volume filling factor increases,
bubbles grow in size, slowed by recombinations.  Finally
bubbles saturate in radii and have similar sizes.  In this
scenario we expect to find small bubbles with a large scatter in size
during early reionization, and large bubbles with similar sizes at
late reionization.  The narrowing of bubble width with increased ionization
fraction was also seen in the studies of limiting cases in earlier work.

For MHR recombinations, the bubble distributions do not narrow at late times,
instead recombinations are so weak that the bubble sizes tend to have
runaway behavior, getting larger and spreading more as reionization proceeds.
In the examples shown,
MIPS recombinations are seen here to limit the growth of bubble regions,
simply because MIPS recombinations are so much larger in number than MHR
recombinations.  However,
in models where more photons are present (large $\bar{x}_i$) 
the radial distribution for MIPS
models
moves to large $R$ as is seen in the MHR case above for late times (large 
$\bar{x}_i$).

 \subsection{Ionization Fraction}
\begin{figure*}[htb]
\begin{center}
\resizebox{5.5in}{!}{\includegraphics{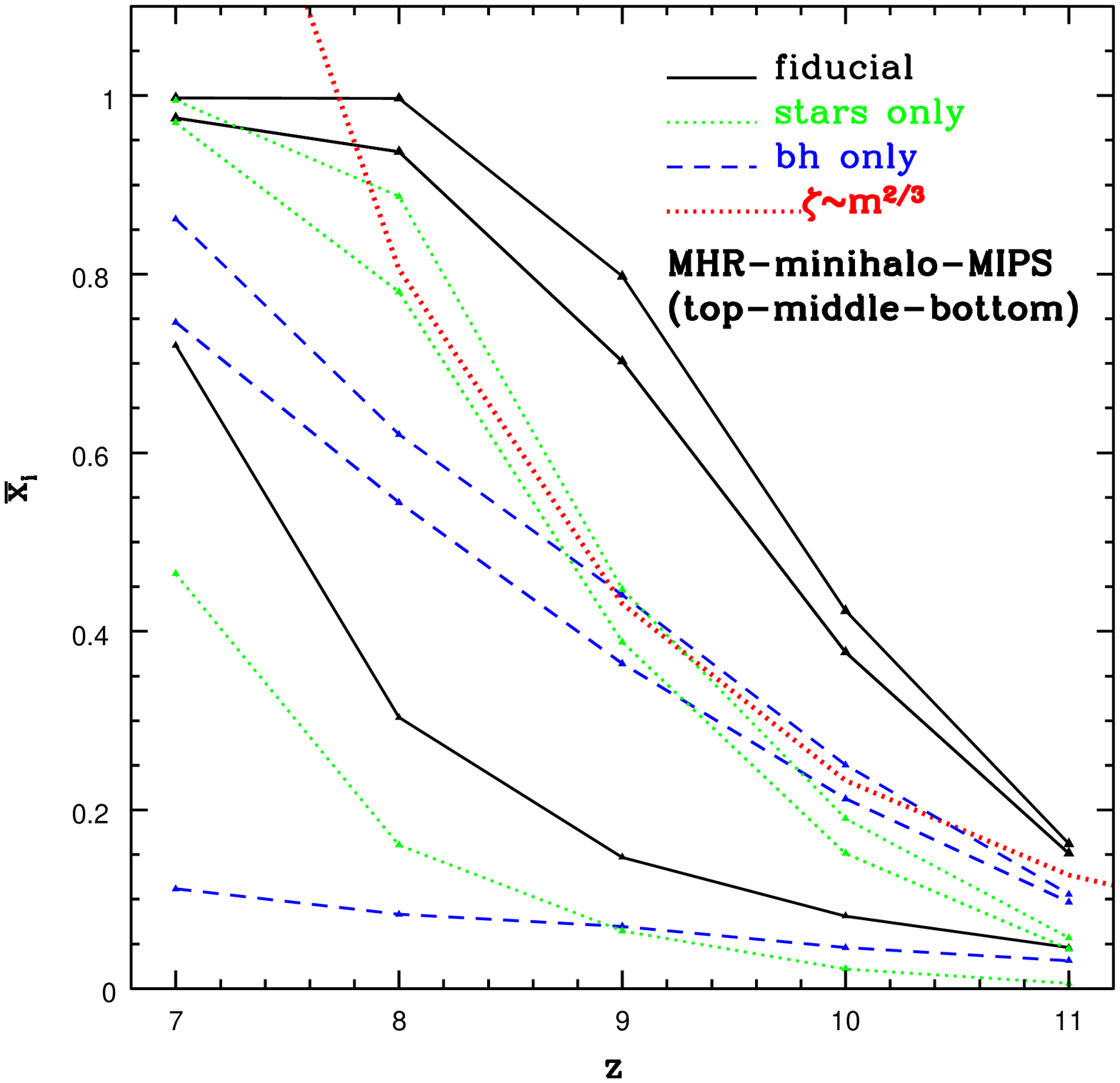}}
\end{center}
\caption{$\bar{x}_i$ as a function of redshift
starting with $\bar{x}_i = 10^{-3}$
at $z=12$.  The different lines of the same
type are for MHR (top), minihalo (middle), and MIPS (bottom) recombination.
The minihalo (center) lines are shown
for two different Monte Carlo runs to illustrate the
run to run scatter.
The solid line is the fiducial model,   
the short dashed and dotted lines correspond to the models with only black
holes and stars respectively.
The heavy dotted line is the ionization fraction for a
$\zeta(m) \sim m^{2/3}$ model (eq. \ref{zfmh}, similar to FMH05).
}
\label{fig:xbarz}
\end{figure*}

The global ionization fraction is the fraction of mass inside bubbles
at any given time; time histories are shown
in Fig. \ref{fig:xbarz} for the MHR, minihalo and MIPS
recombination ansatze with the 
fiducial photon production model.  Even with negligible photons present
initially, the ionization fraction gets close to 1 by $z=7$ for
minihalo and MHR recombinations.  For
comparison, a model similar to FMH05 is also shown (given
by equation \ref{zfmh}).
At early times the
ionization fraction grows more slowly for this time independent $\zeta$ model
as the mergers increase photon production as a function
of mass more steeply than the $m^{2/3}$ slope.
At late times recombinations become more important in the full models and
slow reionization, hence, relative to them
$\bar{x}_i$ for the FMH05 model increases more quickly.
The cases for stars only and $3.5 \sigma$ black holes only are shown as well.
At late times the ionization fraction in the stars only model increases 
relative to the black holes only model,
in part this is due to the rise in
the black hole minimum host halo mass and the stronger decline
in black hole merger photon production with redshift as seen in
Fig. \ref{fig:fourzeta}.

\subsection{Scatter: model variations}
\label{scatter_results}
There is scatter in the results both from model-to-model variations
and from the merger-induced scatter within one model.  The former represents
uncertainties in the modeling, the latter scatter is expected even
if the input parameters are perfectly known.

\begin{figure*}[htb]
\begin{center}
\resizebox{5.5in}{!}{\includegraphics{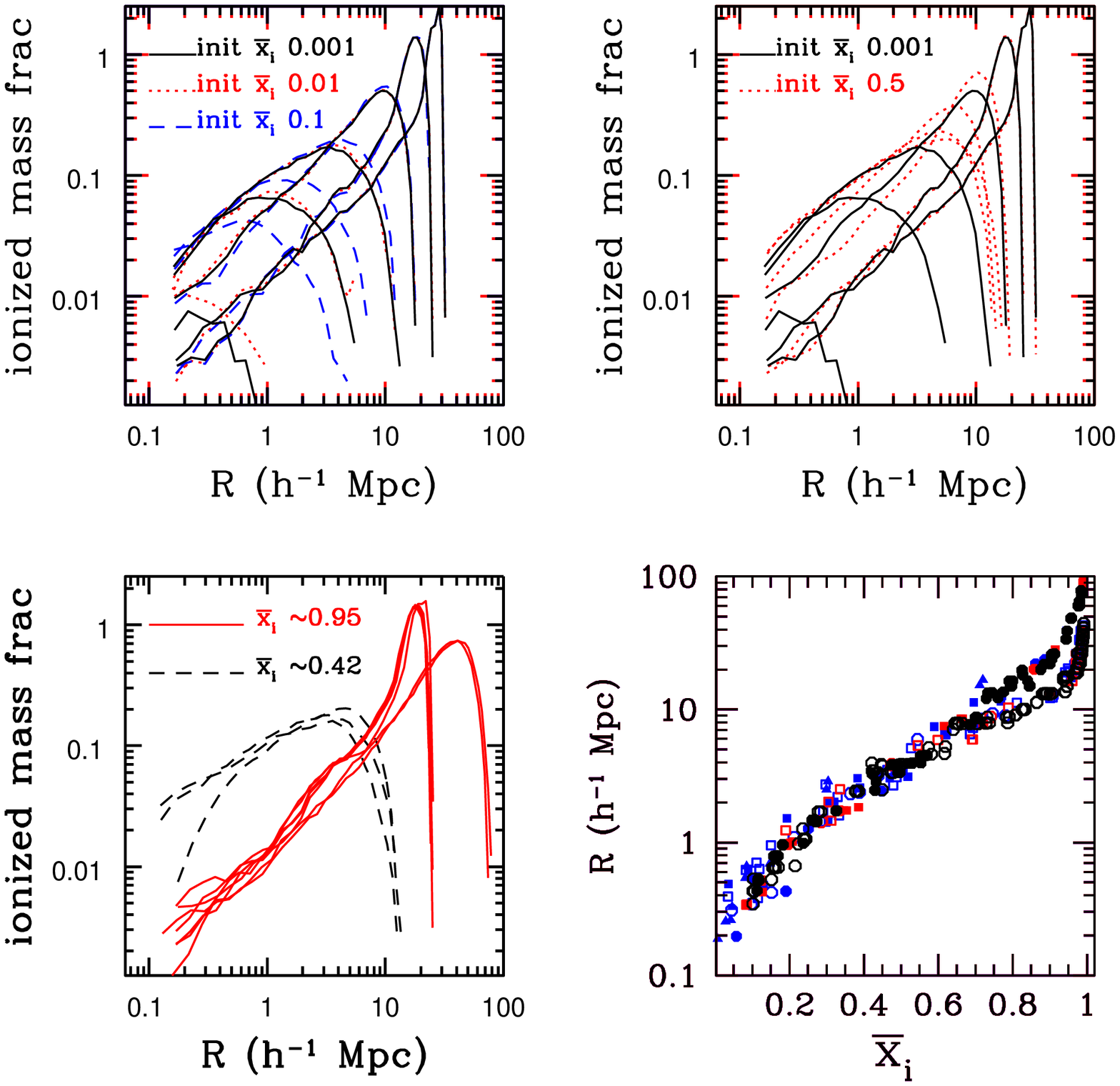}}
\end{center}
\caption{Variations due to different modeling assumptions.
Top left and right: initial condition dependence for minihalo recombination.
At left, initial ($z=12$) 
$\bar{x}_i = 10^{-3}$ (solid),$10^{-2}$ (dots),$10^{-1}$ (dots),
for $z= 11.9, 11, 10, 9, 8, 7$.  At right
initial $z=12$  $\bar{x}_i = 10^{-3}$ (solid),$0.5$ (dots), for the
same redshifts.
Bottom left: 
distributions for
models with $\bar{x}_i \sim$ 0.42 (dashed), 0.95 (solid).  For low $\bar{x}_i$
the $\ln R $ distributions are similar, irrespective of recombination or other 
modeling assumptions, for high $\bar{x}_i$ the profile depends strongly
on the recombination choice (the rightmost peak is MHR recombination, the
peak left of it is minihalo recombination, for several models).
Bottom right: $R_c$ as a function of $\bar{x}_i$ is shown for all
the models to illustrate trends and model to model scatter.  
Open symbols are minihalo recombinations.
Squares are for $n= 1.05$ models, $f_* f_{esc} \rightarrow
4 f_* f_{esc}$, $\Omega_b h^2 =0.0225$, 
octagons are for the fiducial model with varied initial conditions,
including an initial condition set at $z=16$, two variations of $m_{0,min}$
$3 \sigma$ or $3.5 \sigma$ seed black holes, $\alpha = 2/3$ or $\alpha = 0$,
and stars only and $3.5\sigma$
black holes only.  
Filled symbols are 
MHR (octagon/square) recombinations.  Filled triangles are MIPS recombinations,
show for the fiducial model and 4 other cases: the 
variation taking $\alpha =1$, and the other 3 combinations of
$3.5 \sigma$ or $3 \sigma$ seed black holes and
$m_{0.min} = 0.1m$ or $m(T=10^4K)$. }
\label{fig:varyic}
\end{figure*}

In order to see similarities and variations between models, we show
four different comparisons in Fig. \ref{fig:varyic}.
The first model to model
scatter we consider is that due to the unknown initial conditions
for our formalism.
The two top plots show the effects of the $z=12$ initial conditions
on the ionized mass fraction for different redshifts.
The curves are for the fiducial $\bar{x}_i(z=12) = 10^{-3}$ model
and
$\bar{x}_i(z=12) = 10^{-2},10^{-1}.$  At right we show the
fiducial model and $\bar{x}_i(z=12)=0.5$.
By $z = 10$ all the models with $\bar{x}_i$ at $z=12 \leq
0.1$ predicted close to identical bubble radii distributions.
The model with $\bar{x}_i=0.5$ at $z=12$ has the radii
and ionization fraction changing much more slowly between $z=12$ and
$z=9$, and then converging to the other cases.
In addition, the $\bar{x}_i(z=16) = 10^{-3}$ and  
$\bar{x}_i(z=12) = 0.16$ models (sharing the same $\bar{x}_i$ at $z=12$,
not shown above) gave extremely similar 
radial distributions to each other at the redshifts above.

The bottom two plots consider a sampling of different models: different
redshifts, photon production rates, recombinations, cosmologies, etc.
At lower left models with similar $\bar{x}_i$ but otherwise widely varying
assumptions are shown.  
For low global ionization fraction the different models give distributions that
are quite similar, irrespective of recombination method (and other
properties such as redshift or photon production methods).
For high global ionization fraction the distributions
depend much more strongly on recombination methods and not only the global
ionization fraction $\bar{x}_i$. 
The $\bar{x}_i \sim 0.95$ models with MHR recombination peak to larger
$R$ relative to the minihalo recombination models.
The characteristic radius $R_c$ for several hundreds of our model variants is
shown at lower right,
the general trend of larger radii corresponding to larger
ionization fraction is clearly visible, with a spread that can be
read off of the plot.

\subsection{Scatter: merger induced for a fixed model}
The scatter in the photon production rates for a given model due to
mergers was described in
\S\ref{scatter}.  To propagate the scatter in photon
production rates calculated from Eq. (\ref{eq:scatter}) to the scatter
in bubble barriers, we replace 
\eq 
\label{eq:onesig}
\int dt (\zeta f )_t \rightarrow
\int (\zeta f)_t \pm \sqrt{\Delta (\int (\zeta f)_t dt)^2  +\sigma_{ic}^2}
\end{equation}
in Eq. (\ref{eq:barrier}).  We add the scatter of the initial
condition ($\sigma_{ic}^2$, corresponding to
the first term in equation \ref{eq:barrier}) in
quadrature to $\Delta (\int (\zeta f)_t dt)^2$.  
We take this initial condition scatter to also be Poisson in accord with our
assumption (and that of FMH05) that the initial condition
sources are Poisson distributed
in the bubbles.  The resulting barriers (overdensities) are those required 
for a region of mass $M$ to be ionized if the sources within have
the one-sigma fluctuation calculated above.

\begin{figure*}[htb]
\begin{center}
\resizebox{5.5in}{!}{\includegraphics{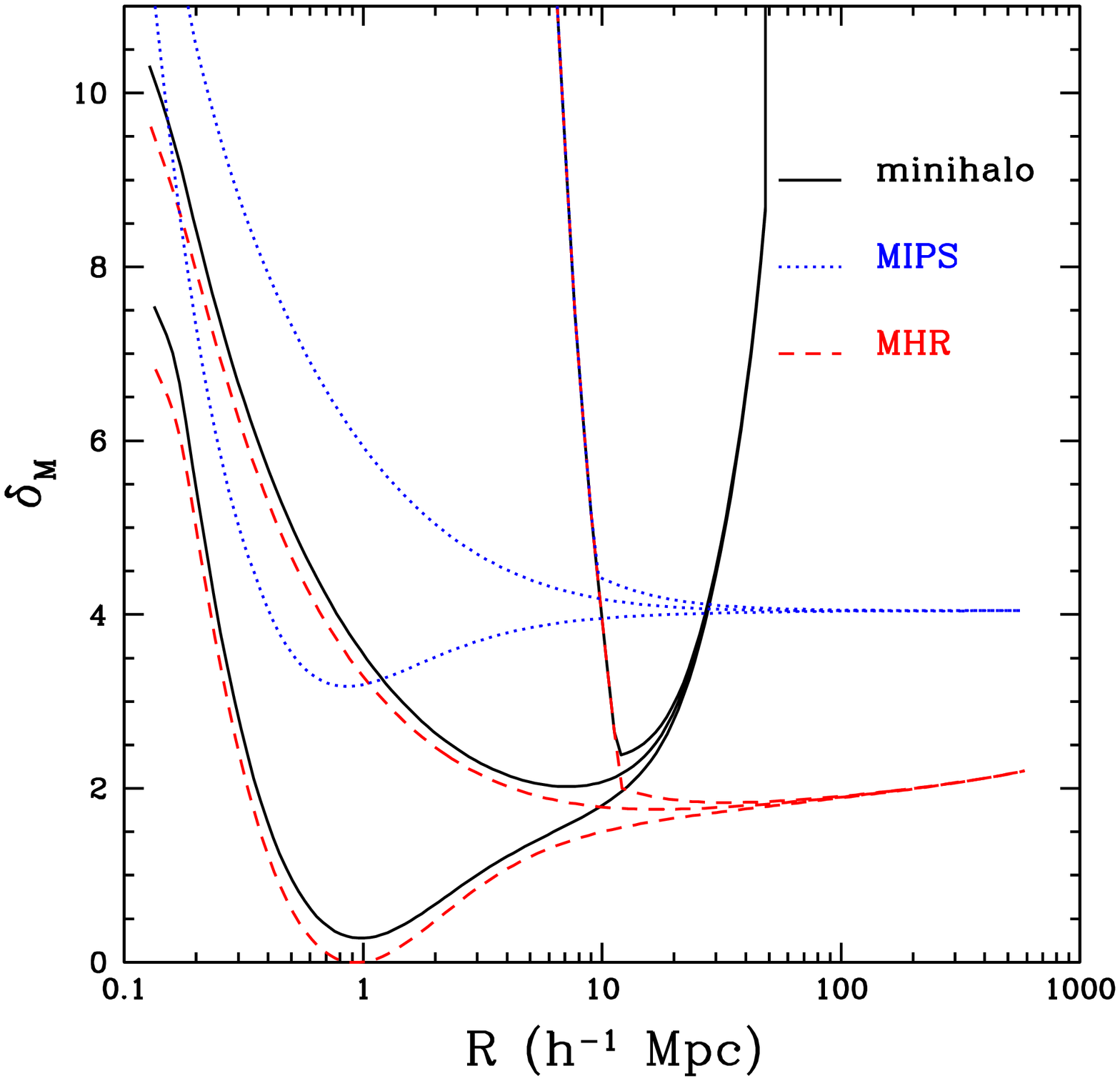}}
\end{center}
\caption{Barriers at $z=10$ with scatter for the fiducial model.
The central line is the mean, lines above
and below are $1 \sigma$ scatter in the time integral of 
$\zeta_{t,a}(m)$ as using the replacement of
equation \ref{eq:onesig}.  The solid/dashed/dotted lines are minihalo,
MHR, MIPS recombination respectively.  The scatter to fewer photons
results in a larger bubble size: a larger volume is needed in order to
enclose a sufficient numbers of sources.}
\label{fig:z10barrier}
\end{figure*}

The corresponding barriers for the fiducial model are shown
at $z=10$ in Fig.~\ref{fig:z10barrier}, for all three recombination models.
The central line indicates the average
barriers $\delta_M$, while the higher (lower) barrier $\delta_{M\pm}$
corresponds to the smaller(larger) photon production rate 
of one sigma fluctuations in $(\zeta f)_t$.
At small bubble radii, the scatters around the mean values
are large, while at the largest radii, the three barriers converge to the
(recombination dependent) limiting bubble radius. 
There are two effects. Smaller bubbles presumably
contain fewer halos, so the scatter of photon rates
per source plays a
relatively important role in determining the required overdensity for
bubbles; conversely, the largest bubbles contain more halos and thus the
effects of the scatter tend to average out within. In addition the
barriers and thus radii for the scattered and not scattered cases converge when
they hit the limiting value set by recombinations.  At later times the
distribution of ionized mass fraction as a function of $R$ becomes more
and more skewed to large radii, causing the characteristic radius to
converge with this limiting radius.
This trend was seen in all the cases where a limiting radius
was found.  
Producing fewer photons can lead to a scatter
up or down in bubble radius: fewer photons means fewer ions all around,
but the bubbles present are often required to
be larger in order to enclose a sufficient number of sources.
As in the case without scatter, for low $R$ MHR, MIPS 
and minihalo recombinations 
are relatively weak and thus give similar results.  

\begin{figure*}[htb]
\begin{center}
\resizebox{5.5in}{!}{\includegraphics{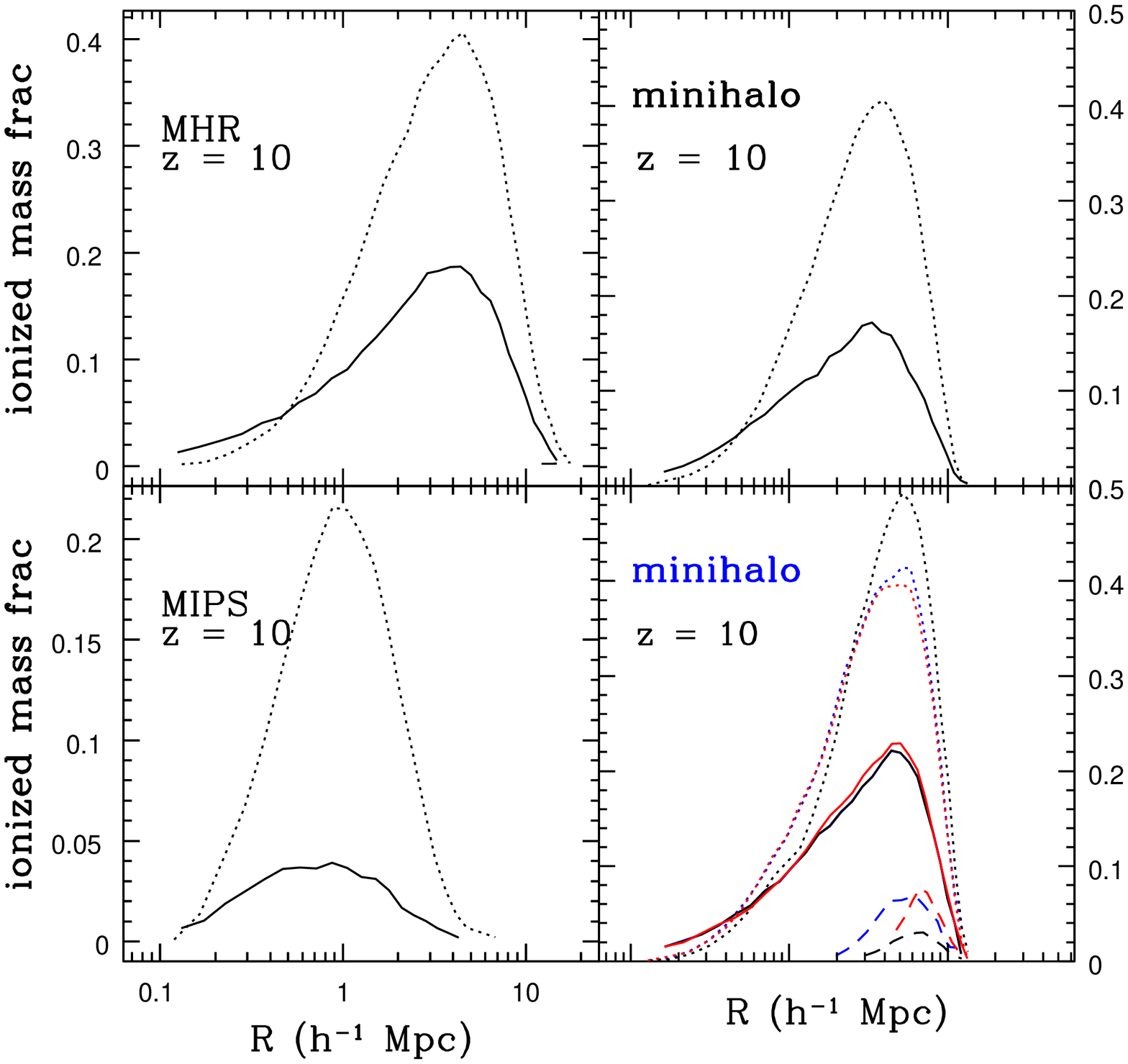}}
\end{center}
\caption{
Top left to right and bottom left:
Radii at z = 10 and their one sigma scatter up and down
for our fiducial case for MHR ($\bar{x}_i = 0.42$), minihalo
($\bar{x}_i = 0.38 $), and MIPS ($\bar{x}_i = 0.08$)
recombination.  Bottom right: 3 estimates for
scatter at z=10 for a model with $\bar{x}_i(z=12) = 0.16$: starting
at $z=16$, starting at $z=12$ with usual
scatter and starting at $z=12$ doubling the usual scatter.  The lines for
the scatter to bigger radii starting at z=16 and that starting at z=12 are
very similar by $z=10$.  The highest dotted line shows how doubling the
scatter translates into a change in the characteristic radii.
Solid lines are the mean, dashed lines are scatter to fewer photons,
dotted are scatter to more photons.  Lines not seen indicate an
ionization fraction too small to appear.
}
\label{fig:compfmh}
\end{figure*}

The corresponding radii distributions for these barriers are shown
at top left and right and bottom left in Fig. \ref{fig:compfmh}. 
At bottom right the scatter is shown for the minihalo case for the
model with $\bar{x}_i = 10^{-3}$ at $z=16$, for the corresponding
fiducial model with $\bar{x}_i = 0.16$ at $z=12$ and for this fiducial
model with the scatter doubled.  The scatter for the model evolved from
$z=16$ converges to that starting at $z=12$ by $z=10$.

The full ionized mass fraction distribution is found by
combining the scattered barriers and the mean one in a distribution,
with some weighting.  This weighting should correspond to outcome of the 
following procedure.  In principle there is a sequence of
barriers corresponding to adding and subtracting the
fluctuations with a continuous coefficient rather than just 
the two cases above, with larger coefficients having smaller weights
in the joint distribution.  Each random walk used to find
the ionized mass fraction then has a different barrier
(``walk barrier'')
sampled from this distribution of barriers.  At large scales all 
the walk barriers
coincide, but at smaller scales fluctuations between
different possible barriers come into play to give a range of
walk barriers.
The first crossings of each walk barrier are dependent both on
the shapes of the barriers in the distribution and
on the scales at which each barrier appears.  A fluctuation from
a higher barrier
down to a lower barrier in a walk barrier at some mass $M$ will produce
first crossings not only of a path which was counted in the 
ionized mass fractions shown above, but also of any path which would have
had first crossings earlier for this lower barrier but didn't for the
higher barrier.  If there was only one such transition between the barriers
above and it was at a fixed $M$ for all walk barriers,
there would be a large spike at the mass scale of the transition,
due to all these paths being included at the same $M$.  (Likewise
a drop would be expected for a sole transition from a low barrier to a high
barrier in a walk barrier.) 
We do not expect these extreme cases to occur for many walk barriers however,
and the transition masses $M$ will change as well.  
We make a rough estimate of the resulting distribution as follows.
Most of the time the walk barrier will sample the different
barriers often enough that a ``pileup'' of paths that would have
crossed earlier, causing such a spike, should be suppressed, similarly
for the sharp signal associated with a single jump up to a higher barrier.
In general, we expect each walk barrier to sample our 3 barriers
frequently, and effects from different walk barriers (different
samplings of the barrier distribution) to average out, so
that the resulting ionized mass fraction
distribution becomes close to a weighted average of the 
corresponding distributions for the 3 barriers shown above.
 
\begin{figure*}[htb]
\begin{center}
\resizebox{5.5in}{!}{\includegraphics{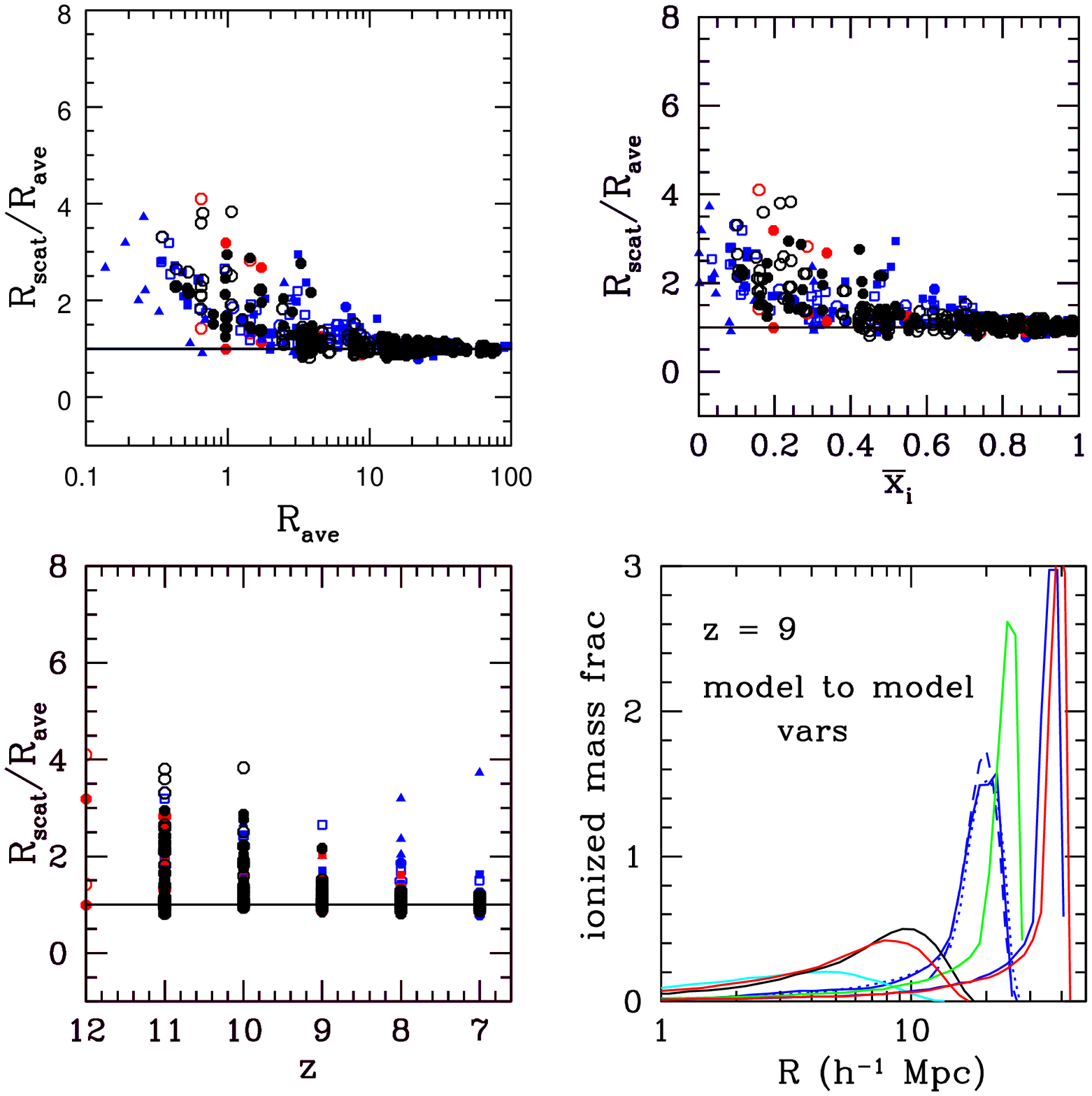}}
\end{center}
\caption{Top left:
The ratio $R_{scat}/R_{ave}$ amongst several models as
a function of $R_{ave}$.  Top right:
The range of the mean value of $R_c$ as a function
of $\bar{x}_i$.
Bottom left: $R_{scat}/R_{ave}$ as a function of redshift.
For all three of these, models with $R_c > 100$ are not shown.
Filled symbols are MHR (octagon/square) and MIPS (triangle) recombinations, 
open symbols are minihalo
recombinations, specific models are as in Fig. \ref{fig:varyic}. 
Bottom right: many different models are shown at $z=9$.  Left to right
(smallest to biggest when there is overlap): fixed $m_{ref}$,
$\Omega_b h^2 = 0.0225$, the fiducial model, $n$ = 1.05,$f_*f_{esc}\rightarrow
f_*f_{esc}*4$ (with scatter), $3 \sigma$ black holes,$\alpha = 0, 3.5 \sigma$
black holes, $\alpha = 0, 3 \sigma$ black holes. 
}
\label{fig:rscat}
\end{figure*}

Some general trends were seen across many models and redshifts.
The scatter effects for the families of models are summarized in
Fig. \ref{fig:rscat} (as mentioned above MHR and MIPS 
cases with $R_c > 100 h^{-1}$ Mpc
are left out but models with $\alpha = 0$ are included). 
The final scatter (z=7) seems small in all other cases and the radii converge
to a value independent of initial ionization fraction (but dependent upon
each model's photon production rates).
We call the peak (characteristic radius)
of the distributions in $R$ (such as above) 
for the scattered model
$R_{scat}$ and the peak for the fiducial model $R_{ave}$ in these plots.
The scatter decreases with increasing radius and decreasing redshift.
This is shown in the upper and lower left plots:
the upper left
plot shows the effect of reduced scatter with increasing numbers of
sources ($R$) included as mentioned earlier, at lower left is the 
redshift dependence.
The ratio $R_{scat}/R_{ave}$ also decreases as a function of increased ionization
fraction $\bar{x}_i$, shown at upper right.  As mentioned above, this is in
part due to the ionization fraction profile becoming limited by recombinations
as the ionization fraction grows, with the characteristic radius converging to
the limiting radius.
At lower right it is shown that the model-to-model scatter is often
larger than the merger-induced scatter.  The average curves for 
$3\frac{M^2}{\bar{\rho}} n_b(M,\delta_M)$, the fraction of total H in
ionized H bubbles with a given $\ln R$ 
are shown for a series of models at $z=9$ along with
the representative scattered curves for one of them.
Note that the mass ratios chosen to
define a major merger will also affect the scatter,
for example relaxing the major merger mass ratio leads to more mergers overall 
and thus smaller Poisson fluctuations.  

\subsection{Bubble Boundaries}
\begin{figure*}[htb]
\begin{center}
\resizebox{5.5in}{!}{\includegraphics{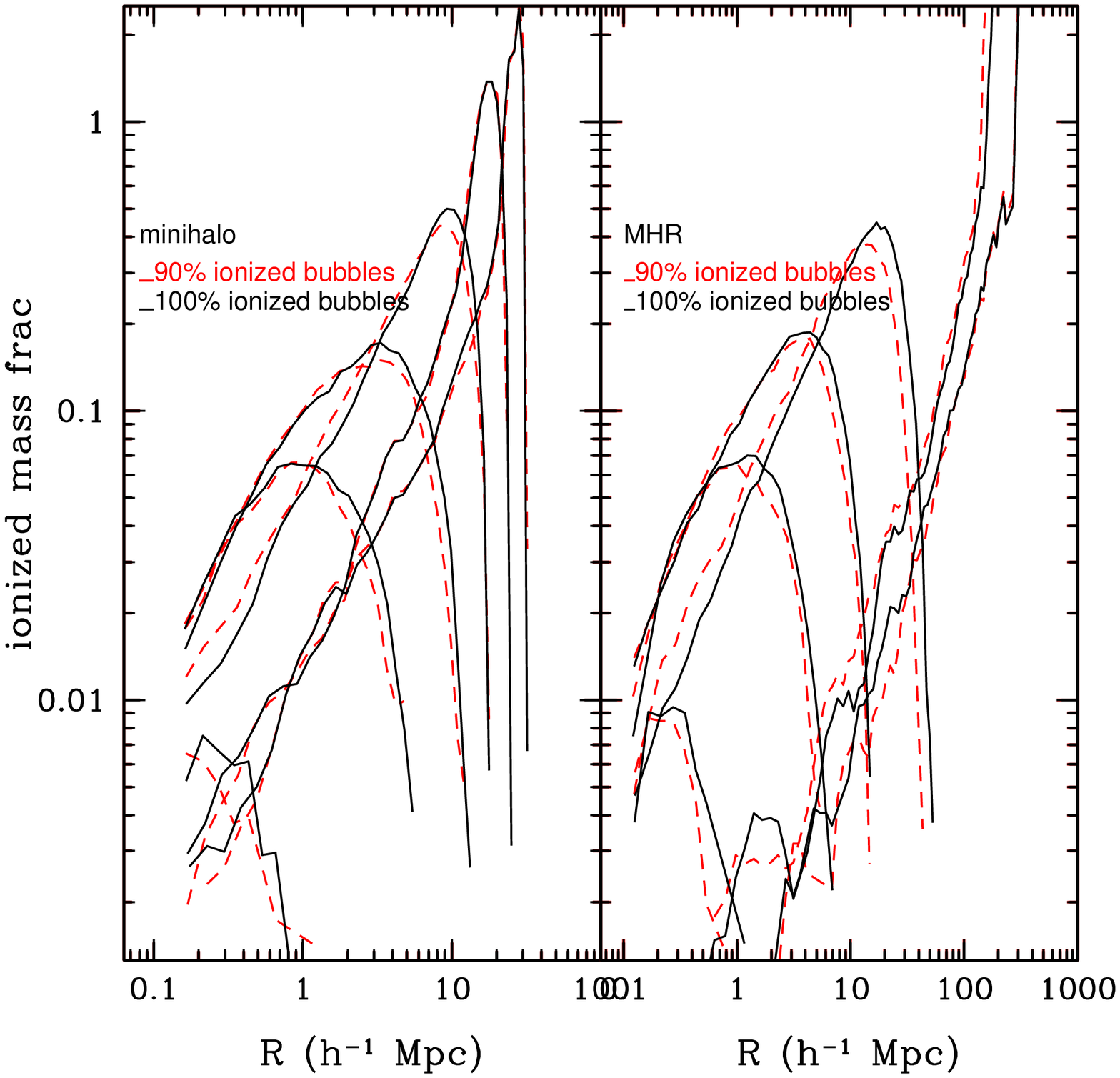}}
\end{center}
\caption{Mass fraction in bubbles as function of radius for 
bubbles which are  100\% ionized (solid) and 90\% ionized (dashed) at
redshifts 11.9, 11, 10, 9, 8, 7 (lowest to highest curves) for
minihalo and MHR recombinations (left and right).  The effects
of changing between fully ionized to 90\% ionized bubbles does
not change the distribution as a function of radius that much except
at large ionization fraction.  Note the radial scale for MHR recombinations is
much larger.
}
\label{fig:evan}
\end{figure*}

One interesting question is
how the ionization fraction decreases from 100\% as one goes
outside a fully ionized bubble at any given time (we thank Evan Scannapieco
for suggesting this calculation and Steve Furlanetto for discussions
about its interpretation).  If the ionization fraction dropped from fully
ionized to e.g. 90\% ionized very slowly in space, 
or if bubbles were very close and connected by
high density bridges, relaxing the constraint from fully ionized to 90\% ionized
would give a very different distribution of barriers and corresponding
radii.  To calculate the 90\% ionized case we
relax the complete ionization requirement for bubbles to
\eq
\begin{array}{l}
\int dm \zeta_{FMH}(m,t_{bc}) P_1(m,t_{bc}|M,\delta_M) 
\\ 
\; \; \; \; +
\int_{t_{bc}}^t dt' [(\zeta f)_{t'} - A_u (1+\delta_{R,M}(t'))
\frac{M_{ion}}{M} C(R_{ion}(t'))] = 0.9 \; .
\label{eq:barrier9}
\end{array}
\end{equation}
The bubble radii, for minihalo and MHR
recombinations, are compared for 90\% ionized and 100\% ionized bubbles
in Fig. \ref{fig:evan}.
From Fig. \ref{fig:evan} it is seen that
these would-be bubbles tend to have very similar sizes compared
to completely ionized bubbles.
The main exception is when the global ionization
fraction is close to one for MHR recombination, 
perhaps indicating that a significant
amount of bubble merging is occurring and/or that there are
large regions near the bubbles (perhaps not in 100\% ionized bubbles)
with high 
average ionization fraction.  For the above models and
a sample of others,
the fraction of mass 90\% ionized or more is within 6\%
of the mass in regions which are fully ionized at $z =7$,
the difference
increases to $< $15\% at z = $11$.

\section{Conclusions}

We have calculated
contributions to $\zeta_{t,a}(m)$ including mergers,
a time and mass dependent phenomena. 
We developed a time dependent description of the requirements
for bubble formation ($\delta_M$) that generalizes previous formulations.  
Using three recombination histories and several different parameterizations
of the uncertain photon production, the requirements
for bubble formation, characteristic radii and ionization fraction
histories were then found.

Our results can be summarized as follows.  The resulting solutions for
$\delta_M$ 
from our formalism and calculations 
interpolate between the two limiting cases considered in previous work
as expected.  In comparison with these cases,
the increased photon production due to mergers gives a faster
rise in ionization fraction at early times, and then a slower one as
recombinations become important.  At early times the characteristic
radii and distribution as a function of bubble radius for ionized mass 
depends most strongly on the global ionization fraction.
At later times, the radial distribution for identical
global ionization fractions depends strongly upon recombination
prescriptions.
The width of the distribution for the bubble
radii narrows for minihalo recombinations at later times just as found in
earlier work.  
For MHR and MIPS recombinations the distribution as a function of
$R$ widens once the ionization fraction is large.  In addition, fixing
global ionization fraction, the
characteristic radius at late times tends to be larger for MHR and MIPS
recombinations relative to minihalo recombinations.
We also estimated the scatter between models (fairly large for
different recombination prescriptions as described above, 
decreasing with time for similar models differing in
initial conditions) and the scatter of radial distributions
of the ionized mass within a given model.  For the latter,
recombinations (except when runaway behavior is present)
limit the size of the largest bubbles independent of whether
there is scatter present or not.  As the global ionization fraction
increases, the characteristic bubble size tends to this limiting size,
resulting in characteristic bubble sizes with very small scatter as well.
The
model-to-model scatter tends to be larger than the merger induced scatter
within a model.

We also explored variations using the WMAP cosmology parameters,
with the seed black hole prescription mentioned earlier.
Because the estimates of the halo masses containing black holes 
are even more unconstrained for this case, and for simplicity, these models 
are not included in the plots above.  Summarizing these cases, 
using the WMAP cosmology parameters and the accompanying family
of models reduces strongly the number of photons, 
mostly because high mass halos are
much rarer at early times.  The WMAP models of radii vs. ionization fraction
overlapped with those seen in Fig. \ref{fig:varyic}--i.e. there was no
additional spread introduced.  The results for recombination dependence
and ionization fraction followed the trends noted above.
The WMAP cosmology with $5\sigma$ seed
black holes gave larger scatter simply because there were fewer sources and
hence the Poisson fluctuations were larger.

We also explored how the ionization fraction changed as the bubbles were
required to not be fully ionized, but only 90\% ionized.  These mostly ionized
bubbles had about the same amount of mass in them as the fully ionized bubbles,
indicating that slightly relaxing the constraint of full ionization doesn't 
change the mass fraction contained in the relevant bubbles drastically.

The formalism and tools developed here and results found can be extended
in future work.
Many other parameter choices also fall well within the range of reasonable
guesses due to the large uncertainties at early times, explorations of
these would be very interesting.   Pop III stars can be included in the
same framework as well.   The observational consequences of the
trends we have identified in our models so far are also a clear next
step.  This can be done using analytic methods based on the distributions
we have found.  Another route would be to
implement the $\zeta_t$ merger
prescriptions into numerical simulations of histories 
or semi-analytic simulations such as those of Zahn et al \cite{Zah06},
and calculate the observational consequences this way.

Our methods to include mergers 
might also be useful to include in other descriptions of reionization,
e.g. those which include the effects of clustering of galaxies
such as is done by Babich \& Loeb \cite{BabLoe05}.

We both thank the Aspen Center for Physics and the January winter meeting
organizers for the opportunity to present this
work when it was near completion.  We are grateful to many people for
helpful suggestions, pointers to literature and 
explanations, including
A. Barth, M. Boylan-Kolchin,
D.Eisenstein, J. Greene, J. Hennawi, G. Howes, I. Iliev, D. Keres,
M. Kuhlen, C.P. Ma, A. Meiksin, C. McKee, 
M. Morales, P. Norberg, E. Quataert, B. Robertson,
E. Scannapieco, R. Sheth, F. van den Bosch, A. Wetzel and O. Zahn.  
We want to thank
S. Furlanetto in particular for numerous explanations and discussions
and both him and E. Scannapieco for comments on the draft.  We are
very grateful to the anonymous referee for many suggestions and corrections that
clarified the work discussed in this paper.
Last but not least, JDC thanks M. White for more
discussions about this topic than he ever wanted to have.
JDC was supported in part by NSF-AST-0205935 and T.C. Chang was supported in
part by NSF as well.

\section*{Appendices}
\subsection*{Extended Press-Schechter definitions}

Two probabilities appear in our formalism in the text.
The probability that a halo of mass $m$ at time $t$ 
contains 
a halo of mass $m_0$ at time $t_i$ is:
\eq
\begin{array}{l}
P_1(m_0,t_i|m,t) dm_0  \\
\; \; =
\sqrt{\frac{1} { 2\pi}} 
|\frac{d \sigma^2(m_0)}{d m_0} |
\frac{\delta_c(t_i) -\delta_c(t)}
{[\sigma^2(m_0) -\sigma^2(m)]^{3/2}}
\exp \{ -\frac{(\delta_c(t_i)-\delta_c(t))^2}{2(\sigma^2(m_0) -\sigma^2(m))}  
\} dm_0
\end{array}
\end{equation}
The overdensity for collapse at time
$t_i$ is taken to be time dependent,
 $\delta_c(t_i) = 1.68/D(z_i)$ where $D(z_i)$ is the growth
factor, and
$\sigma_i^2 = \sigma(m_i)^2$ and $\sigma^2 = \sigma(m_0)^2$, 
the variance of the linear power spectrum
smoothed over a region of mass $m$ (at z=0).

The other quantity is the time derivative of the above,
the fraction of halos that have mass $m_0$ that 
have jumped at time $t_i$ from mass
$m_i$: 
\eq
\dot{P}_1(m_i\to m_0;t_i)  \frac{m_0}{m_i} dm_i dt_i =
\frac{1}{\sqrt{2 \pi}}
\frac{1}{(\sigma_i^2 - \sigma^2)^{3/2}}
[-\frac{d \delta_c(t_i)}{dt_i} ] |\frac{d \sigma_i^2}{d m_i}|
 \frac{m_0}{m_i}
dm_i dt_i \; .
\end{equation}
The factor of $m_0/m_i$ is added to convert from
the number of points originating from $m_i$ halos to the number of
points in $m_0$ halos containing these earlier $m_i$ halos.

To write the merger fraction, Eq.\ref{mergeprob}, consider the following.
The number of merged halos is
the fraction of recently merged halos of mass $m_0$ at $t_i$ 
times the number (density) $n(m_0,t_i)$ of $m_0,t_i$
halos. Multiplying by
their survival probability $P_2(m,t|m_0,t_i)$ of surviving to 
$m,t$ gives the number (density) of $m$ halos at time $t$
which had a merger to mass $m_0 < m$ at this earlier time.
Dividing the number of recently merged halos by the total number (density),
$n(m,t)$ gives the
recently merged fraction.  We then multiply the fraction of halos which
has a recent merger by the fraction of the total number of $m,t$ halos
found in a bubble of mass $M,\delta_M$, i.e. by
$n_h(m,t|M,\delta_M) =
\frac{\bar{\rho}}{m}P_1(m,t|M,\delta_M)$ to give a number density.
The product of these probabilities is then
\begin{eqnarray}
\dot{P}_1(m_i\to m_0;t_i) \frac{m_0}{m_i} n(m_0,t_i) P_2(m,t|m_0,t_i) \frac{1}{n(m,t)}
\frac{\bar{\rho}}{m}P_1(m,t|M,\delta_M)
\end{eqnarray}
which equals Eq. \ref{mergeprob} because
\begin{eqnarray}
P_1(m_0,t_i|m,t) n(m,t) 
\frac{m}{\bar{\rho}}dm_0 dm = P_2(m, t|m_0,t_i) 
n(m_0,t_i)\frac{m_0}{\bar{\rho}} dm_0 dm
\end{eqnarray}
(also this can be used to get an explicit expression for $P_2$).
These probabilities are being multiplied together assuming that 
they are independent.   There is the expectation that overdense regions
should have larger merger rates (e.g. Scannapieco \& Thacker \cite{ScaTha03},
Furlanetto \& Kamionkowski \cite{FurKam05}), however the overdensities here
are in fact quite small (the bubble overdensities $\delta_M$
tend to be between 0 and 4 with corresponding
physical overdensities $\delta_M D(z)$), so that
this effect is expected to be small.

\subsection*{Minimum black hole masses}
We describe here how we chose our minimum host halo masses for harboring black 
holes.  For $m_{bh,min}(z)$ many suggestions of high redshift
black hole histories are available, see for
example the review by Haiman \& Quataert \cite{HaiQua04} and references
therein.  We take our black holes to
be descendants of the very first stars, expected to
be extremely massive due to the difficulty of fragmentation (see for
example Bromm, Coppi \& Larson \cite{BroCopLar99,BroCopLar02},
Nakamura \& Umemura \cite{NakUme99,NakUme01}, 
Abel, Bryan \& Norman \cite{AbeBryNor00,AbeBryNor02},
Schneider et al \cite{Sch02}).  At the end of the lifetimes of these stars,
very massive black holes are expected to form, e.g. Madau \& Rees 
\cite{MadRee01},
however the initial mass distribution of the stars and their resulting black 
holes are unknown (many different examples are considered in e.g.
Alvarez, Bromm \& Shapiro \cite{AlvBroSha05}, O'Shea et al \cite{OSh05},
Scannapieco et al \cite{Sca05}, and Madau et al \cite{Mad04}). 

We follow Madau et al \cite{Mad04} and consider two
cases, putting very massive black holes
in all $>3 \sigma$ or $>3.5 \sigma$ fluctuations at redshift $z=24$,
corresponding to masses $m_{3 \sigma}, m_{3.5\sigma}
(m>1.4 \times 10^5 h^{-1} M_\odot, m>1.4 \times 10^6 h^{-1} M_\odot)$
which then grow primarily through mergers.
We find the minimum mass for black holes at our (later) time of interest 
by requiring at least 90\% of the halos with mass $m_{bh,min}(z)$ to
have at least one halo of mass $m_{bh,min}(z=24)$ in it, 
using extended Press-Schechter.  
(In practice it appears that at least 4 or 5 mergers occur for these high 
$\sigma$ peaks by
$z \sim 15$, we thus require at least the corresponding number of
paths originally at mass $m_1$ to be present in each final mass halo of mass $m$
at time $t$ in order for that mass to host a black hole. We assume
the black holes merge when their halos do, e.g. Mayer et al \cite{May06},
there is still discussion on this issue, see e.g. Madau et al \cite{Mad04}
for further discussion including considerations of black hole ejection.)
A second option for estimating halo masses which can contain
seed black holes is using the requirement of low
angular momentum disks which can collapse, 
the two reference models for minimum black hole
mass used by Koushiappas, Bullock \& Dekel \cite{KouBulDek04}
at $z=9,12,15$ are approximately the
same as those we find extrapolating the minimum mass halos of Madau
et al \cite{Mad04} for $3 \sigma$ and $3.5 \sigma$ host halos at $z=24$.
(For some reason Madau et al \cite{Mad04} get a much lower merger rate than 
we do, ours is more in accord with the calculations of Koushiappas et al 
\cite{KouBulDek04}.)  As we are interested in some reasonable black hole
mass estimates, the similarity between these two calculations suggests
these are useful reference masses to consider.
This is certainly not the only way to estimate the black hole host masses,
other options include tracing the luminosity function back in time e.g.
Wyithe \& Loeb \cite{WyiLoe02}.

Black hole radiation in principle constrains black hole masses and
numbers.  However, compounding the
the black hole/host halo mass relation uncertainties and
accrection mode/speed uncertainties mentioned earlier are 
even more uncertainties tied to black hole radiation estimates.
Harder X-ray photons may
induce additional ionizations (some estimates have X-rays dominate the
luminosity, e.g. Madau et al \cite{Mad04}) in regions with low ionization
fractions, in addition, efficiencies might be higher
(Haiman \& Quataert \cite{HaiQua04}).  
The strongest direct constraints on early black holes is
that they don't produce photons that exceed 
the unresolved soft X-ray background (Dijkstra, Haiman \&
Loeb \cite{DijHaiLoe04}, Salvaterra, Ferrara \& Haardt 
\cite{SalHaaFer05}). This translates into a constraint
on the ratio $\Phi$ of power law to multicolor disk luminosities;
although we use Madau et al's \cite{Mad04} 
original model of including black holes we take
$\Phi < 1$ to avoid the constraints their models 
could not satisfy ($\Phi \sim 0.1$ is common today).   Other
black hole halo mass
constraints requiring more evolutionary assumptions over longer periods
of time include predicting correctly black hole to halo mass ratios today 
(e.g. Ferrarese \& Merritt \cite{FerMer00}), the number
of intermediate mass black holes today
(Yu \& Tremaine \cite{YuTre02}) and the observed quasar activity
observed up to $z \sim 6$.  In addition, one expects starbursts
and black hole growth can be related using the
relation between central black hole 
mass and 
bulge velocity dispersion \footnote{We thank Joe Hennawi
for this suggestion.}, models proposed
to do this explicitly include those by Cattaneo
et al \cite{Cat05} and Enoki et al \cite{Eno03}.
\subsection*{Other scatter assumptions}
Other assumptions are possible for 
\eq
\langle n(m,t|M,\delta_M) n(m',t'|M,\delta_M) \rangle \; .
\end{equation}
and
\eq
\langle n^a(m_0|m,t) n^b(m_0|m,t) \rangle \; .
\end{equation}
besides those used in the main body of this paper.
We list two more possibilities for each.

For
$
\langle n(m,t|M,\delta_M) n(m',t'|M,\delta_M) \rangle$,
one might expect that a large fluctuation at time
$t$ implies a large fluctuation at time $t' = t+\epsilon$,
giving a correlation between different $t,t'$.  In addition,
halos of mass $m$ at time $t$ 
will have a different mass $m'$ at time $t'$, so the fluctuations will
be related for $m \ne m'$ and $t \ne t'$.  

One possibility is to give the scatter a decay time (or characteristic
 $\delta_{rel}$)
and a decay mass $m_{rel}$,
\eq
\langle n(m,t|M,\delta_M) n(m',t'|M,\delta_M) \rangle 
= e^{-(\delta(t')-\delta(t))/\delta_{rel}} e^{-(\sigma^2(m)-\sigma^2(m'))/\sigma^2(m_{rel})} n(m,t)
\end{equation}
Other possibilities are to have the decay depend on $t$ rather than $\delta(t)$
and $m$ rather than $\sigma^2(m)$.

Another possibility is to say that the relation between the mean
values of $n(m,t|M,\delta_M)$ and $n(m',t'|M,\delta_M)$
\eq
\begin{array}{l}
 n(m,t|M,\delta_M) n(m',t'|M,\delta_M)  =\\
 n(m,t|M,\delta_M) \int dm'' P_1(m',t'|m'',t)
\frac{m"}{m'} n(m",t|M,\delta_M ) \; \; \\
m' < m,t'<t \\
\end{array}
\end{equation}
suggests a relation between the scatters for $m'< m,t' < t$
as follows:
\eq
\begin{array}{ll}
\langle n(m,t|M,\delta_M) n(m',t'|M,\delta_M) \rangle &=
\langle n(m,t|M,\delta_M) \int dm'' P_1(m',t'|m'',t)
\frac{m"}{m'} n(m",t|m,t ) \rangle \\
& =
n(m,t|M,\delta_M) n(m',t'|M,\delta_M) \\ &+ 
\frac{1}{V_M}\Theta(t-t')\int dm'' \frac{m''}{m'}
P_1(m',t'|m'',t) \delta(m-m'')n(m,t|M,\delta_M)  \\
&= n(m,t|M,\delta_M) n(m',t'|M,\delta_M) \\ &+ 
\frac{1}{V_M}\Theta(t-t')\ \frac{m}{m'}
P_1(m',t'|m,t) n(m,t|M,\delta_M)  \; ; \; \; \\
& m' < m, t' < t \; . \\
\end{array}
\end{equation}
and similarly when
 $(m,t) \leftrightarrow (m',t')$ when $m'>m, t'> t$.
This is not required: the relation between the average values does
not have to be followed by the scatter, it is an additional
modeling assumption.  

The equal mass and time limit, $m \to m', t \to t'$ can be
taken in the above.  Writing
$\sigma^2(m') - \sigma^2(m) = x,
\delta(t') -\delta(t) = y$ we have
\eq
P_1(x,y) = \frac{1}{\sqrt{2 \pi}} \frac{y}{x^{3/2}}
e^{-y^2/(2 x)} \frac{dx}{dm} = -\frac{d}{dy} 
[\frac{1}{\sqrt{2 \pi}} \frac{1}{x^{1/2}}
e^{-y^2/(2 x)} ] \frac{dx}{dm} \; ,
\end{equation}
then $m \to m'$ is $x \to 0$ and $t \to t'$ is $y \to 0$.
The two limits do not commute, thus another assumption is involved,
we take $m \to m'$ first, roughly writing
$\int dm = \int dm (\Theta(m-m') + \Theta(m'-m) +(\delta_D(m-m') -1))$
and then recognize the Dirac delta function as the $x \to 0$ limit
of the quantity in square brackets.
This also assumes taking this limit and taking the derivative commute.
We then get the full two point function for these assumptions to be:
\eq
\begin{array}{l}
\langle n(m,t|M,\delta_M) n(m',t'|M,\delta_M) \rangle =
n(m,t|M,\delta_M) n(m',t'|M,\delta_M) \\ 
+
\frac{1}{V_M}\Theta(t-t') \frac{m}{m'}
P_1(m',t'|m,t) n(m,t|M,\delta_M)   \\
+
\frac{1}{V_M}\Theta(t'-t) \frac{m'}{m}
P_1(m,t|m',t') n(m',t'|M,\delta_M)  \\
-\frac{1}{V_M} \delta(m-m') \delta_D(\delta(t)-\delta(t')) n(m',t'|M,\delta_M)
\end{array}
\end{equation}
The derivative of
$\delta_D(y)$ is evaluated by integrating by parts and changing variables
to $t,t'$.  

There is also a range of possibilities for the scatter in $\zeta_{t,a}(m)$
or $n^a(m_0|m,t)$.  Again, we describe two.

One possibility is to take the same scatter as is used for
$n(m,t|M,\delta_M)$ and use it for $P_1(m_0,t_i|m,t) =
\frac{m_0}{\bar{\rho}} n(m_0,t_i|m,t)$ and then
define the scatter for $\dot{P}_1(m_i \to m_0,t_i)$ as the limit.
Naive attempts to take this limit give a divergence however.

One also might say that a halo of mass $m_0$ in a halo of
mass $m$ has either had or not had a recent merger,
which suggests a binomial distribution, and that
if the total number of halos of mass $m_0$ in $m,t$ halos is $N$
that this number is then $pN$, implying a scatter
$Np - p^2 N$ rather than the poisson $N$.
Implementing this appears to require yet another assumption because
the time integral of the rate appears, but it will decrease the
scatter from Poisson ($\sim Np$).

None of these scatter prescriptions is required given current knowledge,
thus in the main body of the text we used the simplest version.
It is reasonable to expect that the trends of largest scatter for smallest
radii will continue as large numbers of enclosed halos will tend to 
enclose the mean number of mergers.

\end{document}